\documentclass[manuscript]{acmart}

\AtBeginDocument{%
  }

\usepackage[table,xcdraw]{xcolor}
\usepackage{makecell}
\usepackage{csquotes}

\setcopyright{acmlicensed}
\copyrightyear{2026}
\acmYear{2026}
\acmDOI{XXXXXXX.XXXXXXX}

\acmConference[XXX '26]{Manuscrupt currently under review}{June 03--05,
  2018}{Woodstock, NY}

\acmISBN{978-1-4503-XXXX-X/2018/06}

\def\eg{\emph{e.g., }} 

\def\ie{\emph{i.e., }}

\newcommand{\pquotes}[1]{\textcolor[gray]{0.3}{\textit{#1}}}

\definecolor{lightpurple}{rgb}{0.9,0.8,1.0}
\definecolor{lightyellow}{HTML}{f6f6cf}
\definecolor{lightgreen}{HTML}{cfe4b3}
\definecolor{lightorange}{HTML}{ffe3c8}
\definecolor{lightblue}{HTML}{c2cae4} 
\definecolor{lightpink}{HTML}{fecbe5}  
\definecolor{lightcyan}{HTML}{c9ebe7}

\definecolor{green}{HTML}{D0EBAD}

\begin{document}

\title{ELLA: Generative AI-Powered Social Robots for Early Language Development at Home }

\author{Victor Nikhil Antony}
\authornote{Both authors contributed equally to this research.}
\affiliation{%
  \institution{Johns Hopkins University}
  \city{Baltimore}
  \state{Maryland}
  \country{USA}
}

\author{Shiye Cao}
\authornotemark[1]
\affiliation{%
  \institution{Johns Hopkins University}
  \city{Baltimore}
  \state{Maryland}
  \country{USA}
}

\author{Shuning Wang}
\affiliation{%
  \institution{Johns Hopkins University}
  \city{Baltimore}
  \state{Maryland}
  \country{USA}
}

\author{Chien-Ming Huang}
\affiliation{%
  \institution{Johns Hopkins University}
  \city{Baltimore}
  \state{Maryland}
  \country{USA}
}


\begin{abstract}
Early language development shapes children’s later literacy and learning, yet many families have limited access to scalable, high-quality support at home. Recent advances in generative AI make it possible for social robots to move beyond scripted interactions and engage children in adaptive, conversational activities, but it remains unclear how to design such systems for pre-schoolers and how children engage with them over time in the home. We present ELLA (Early Language Learning Agent), an autonomous, LLM-powered social robot that supports early language development through interactive storytelling, parent-selected language targets, and scaffolded dialogue. Using a multi-phased, human-centered process, we interviewed parents (n=7) and educators (n=5) and iteratively refined ELLA through twelve in-home design workshops. We then deployed ELLA with ten children for eight days. We report design insights from in-home workshops, characterize children’s engagement and behaviors during deployment, and distill design implications for generative AI–powered social robots supporting early language learning at home.
\end{abstract}





\maketitle

\section{Introduction}

\begin{figure*}
  \includegraphics[width=\textwidth]{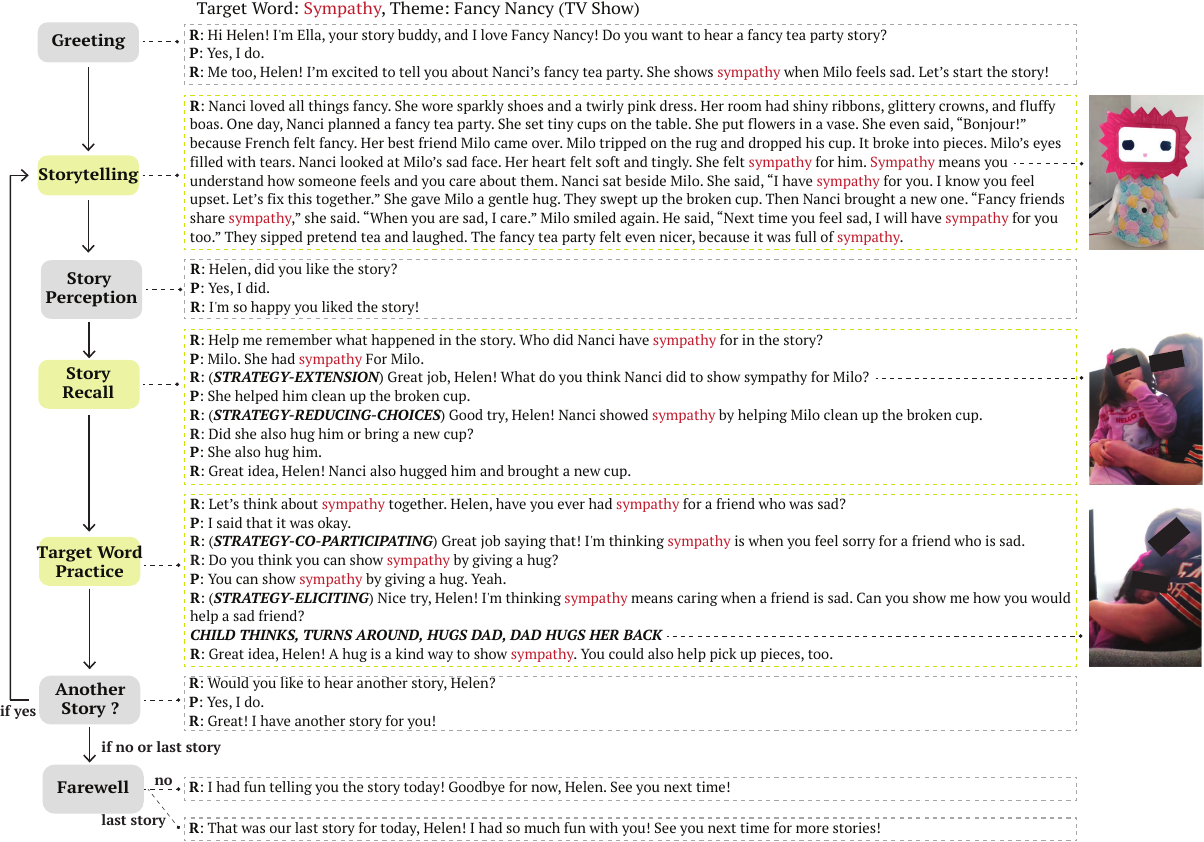}
  \caption{We present ELLA (Early Language Learning Agent), an interactive storytelling robot designed with educators and families to support early language development in children ages 4--6. The robot tells at most four stories per day upon request with a goal of teaching four target words per day determined by parents. The robot incorporates the target words into stories generated using LLMs based on the children's interests and engages the child through targeted and scaffolded interactions. This figure illustrates one sample story session from our deployment study. The children's name was replaced with pseudonyms. The target vocabulary word that the robot is trying to teach is ``Sympathy'' and the theme of this story is the TV show ``Fancy Nancy''. Greeting/farewell from different story sessions were presented to showcase different possibilities.}
  \Description{This figure presents a flowchart illustrating the structure of an interactive storytelling session with ELLA, the Early Language Learning Agent. The target vocabulary word (sympathy) and the story theme (Fancy Nancy, a television show) are specified at the start of the session.
  The interaction begins with a Greeting stage, where the robot introduces itself and asks the child whether they would like to hear a story. If the child agrees, the robot proceeds to Storytelling, narrating a story that embeds the target word in context and provides a child-friendly definition.
  The session then moves through Story Perception, Story Recall, and Target Word Practice stages, during which the robot checks the child’s understanding and encourages reflection on the target word through guided questions.
  At the end of the session, the robot asks whether the child would like to hear another story. If the child chooses “yes,” the flow returns to storytelling for an additional story. If the child chooses “no,” or if the maximum number of stories has been reached, the interaction ends with a Farewell stage.}
  \label{fig:teaser}
\end{figure*}
Early literacy lays the foundation for children’s academic success \cite{dale2023prediction}. Vocabulary knowledge, in particular, is a critical predictor of reading comprehension and underpins children’s ability to communicate, understand, and reason \cite{masrai2019vocabulary}. Learning to read enables children to later read to learn, unlocking access to new concepts, ideas, and forms of knowledge \cite{purpura2011early}. Family engagement plays a central role in children's literacy and language development, as context-rich exposure to language---often occurring through everyday parent–child interactions such as shared reading and informal conversation---supports vocabulary learning and comprehension. Because early language learning is deeply embedded in these daily family routines and interactions, the home is a critical context for children's early language development. However, significant disparities in language and vocabulary development emerge early in life, driven by socio-economic and environmental factors \cite{becker2011social, rodriguez2011trajectories}. The widely cited “\textit{word gap}” illustrates this disparity: by age four, children from lower-income families are exposed to as many as 30 million fewer words than their higher-income peers \cite{neuman2014magic}. Developmental differences and limited caregiver time can further constrain opportunities for rich language interaction at home \cite{mol2014sharing}. While pedagogically informed early vocabulary interventions can mitigate these disparities, such interventions typically rely on sustained human involvement, making them difficult to scale equitably \cite{marulis2010effects, neuman2014magic}.

Social robots offer a promising venue for delivering early language and literacy supports at scale. Child–robot interaction research has long explored robots as tutors, companions, and learning partners, showing that embodied behaviors (\eg emotive speech, expressive gesture) can engage young learners and support vocabulary use. Yet, despite these encouraging findings, much prior work has depended on pre-scripted interactions or Wizard-of-Oz control. Moreover, most studies have taken place in laboratories or classrooms, leaving open questions about how autonomous social robots might function in children’s homes, especially with pre-school–aged children, over longer periods of time.

Advances in generative AI have enabled social robots to move beyond scripted dialogue toward open-ended, adaptive conversation by generating content on the fly, responding to input, and adjusting output in real time \cite{mahadevan2024generative, antony2025xpress, cao2025interruption}. This shift makes it feasible to deliver pedagogically targeted language interventions at home autonomously. Yet generative AI-powered social robots introduce open questions: how to design developmentally appropriate interactions for pre-schoolers to support early language development at-home (\textbf{RQ1}) and how children engage and learn with a generative AI-driven robot over an extended period, at-home (\textbf{RQ2}).

To explore these questions, we followed a three-phased, human-centered design process to develop a social robot for early language development in children aged 4--6 at home. We began by interviewing a diverse set of parents ($n=7$) and educators ($n=5$) to understand existing language-learning practices, needs, and concerns. Drawing on insights from these interviews and prior literature, we developed a generative AI-powered social robot prototype and iteratively refined its design through in-home workshops ($n=12$) with children and their families. This process culminated in the design of ELLA (\textit{Early Language Learning Agent}), a generative AI-powered social robot that supports personalized and targeted in-home early language development for children ages 4--6 through interactive storytelling (see Fig. \ref{fig:teaser}). We tailored ELLA to teach four target words specified by parents through interactive stories with personalized parent-selected themes, and deployed ELLA in the homes of ten children for eight days to examine how children engage with a generative AI–powered social robot, how language learning unfolds during these interactions, and what design opportunities emerge during longer-term, at-home use. This work makes the following contributions:

\begin{itemize}
    \item \textbf{System}: An open-source generative AI-powered social robot (ELLA)\footnote{Supplemental materials (contains prompts used by ELLA and additional experimental details): \href{https://intuitivecomputing.github.io/publications/2026-antony-ella-supp.pdf}{https://intuitivecomputing.github.io/publications/2026-antony-ella-supp.pdf}} that supports in-home early language development for children ages 4--6 through interactive storytelling and an open-source generative AI-powered pipeline for generating personalized and targeted educational content (stories and interactions) for the robot, along with a no-code user interface.
    \item \textbf{Empirical}: Initial empirical evidence demonstrating the efficacy of our social robots in supporting early language development in children ages 4--6 at home. 
    \item \textbf{Design}: Design implications for generative AI-powered social robots for early language learning at home derived through interviews, workshops, and deployment. 
\end{itemize}


\section{Related Works}

\subsection{Early Childhood Language Development}
\label{sec:related-work-early-language-development}

Early language development lays the foundation for children’s later literacy, academic achievement, and communication across domains. A rich home literacy environment plays a central role in this development. Family engagement and everyday interactions, such as shared reading, conversation, and storytelling, provide critical opportunities for children to encounter and use language meaningfully \cite{quintanilla2025family}. Importantly, effective early language development does not arise from passive exposure alone; intentional instructional strategies, including repeated and contextualized word use, significantly enhance vocabulary acquisition \cite{biemiller2006effective, neuman2014magic, silverman2007comparison, beck2007increasing}. For example, children learn words more effectively when they are embedded in interactive, socially grounded activities like parent–child co-reading or conversation \cite{weisleder2013talking}, whereas passive media consumption (\eg television) offers limited benefit for word learning \cite{karani2022influence}.

Access to rich language development experiences, however, is uneven. Socioeconomic disparities in early home environments contribute to substantial vocabulary gaps that emerge as early as age four \cite{hart2003early, neuman2014magic, rodriguez2011trajectories}. Children from higher-income families are exposed to as many as 30 million more words by age three than their lower-income peers; this \textit{``word gap''} has long-lasting consequences for academic outcomes \cite{neuman2014magic}. Children who enter school with stronger vocabularies tend to perform better across subjects—including reading, mathematics, and science—throughout their academic trajectories \cite{purpura2011early, dale2023prediction}. The preschool years are also a period of particularly rapid vocabulary growth \cite{farkas2004detailed}, making early intervention both timely and impactful. 

Daily read-alouds are among the most effective strategies for supporting early literacy, introducing new words, strengthening listening comprehension, and fostering positive attitudes toward reading. Expressive delivery and active participation, such as asking children to recall events, predict outcomes, or retell stories, further deepen engagement and comprehension. Oral language skills developed through these interactions form a critical foundation for later reading and higher-order thinking \cite{li2024research}. Teachers in research-based vocabulary interventions typically introduce $3$--$17$ words per week (6 words on average) \cite{kong2023vocabulary} for pre-school aged children. Most of these interventions use storybooks to teach vocabulary, bringing in $3$--$9$ target words per book or session \cite{kong2024designing}. Instruction sessions can last $5$--$45$ minutes (averaging $20$--$25$ minutes), with frequencies ranging from two to six sessions per week \cite{kong2024designing}.

Building on evidence from prior early language interventions, we designed and deployed an interactive storytelling robot for at-home use. In an eight-day deployment, the robot aimed to teach four parent-selected target words through short (around 5--6 minutes), flexible sessions (up to four per day), while encouraging oral language use and story comprehension practice. We aim to study the potential of generative AI–powered social robots to support early language development at home.

\begin{figure}[t]
  \includegraphics[width=\textwidth]{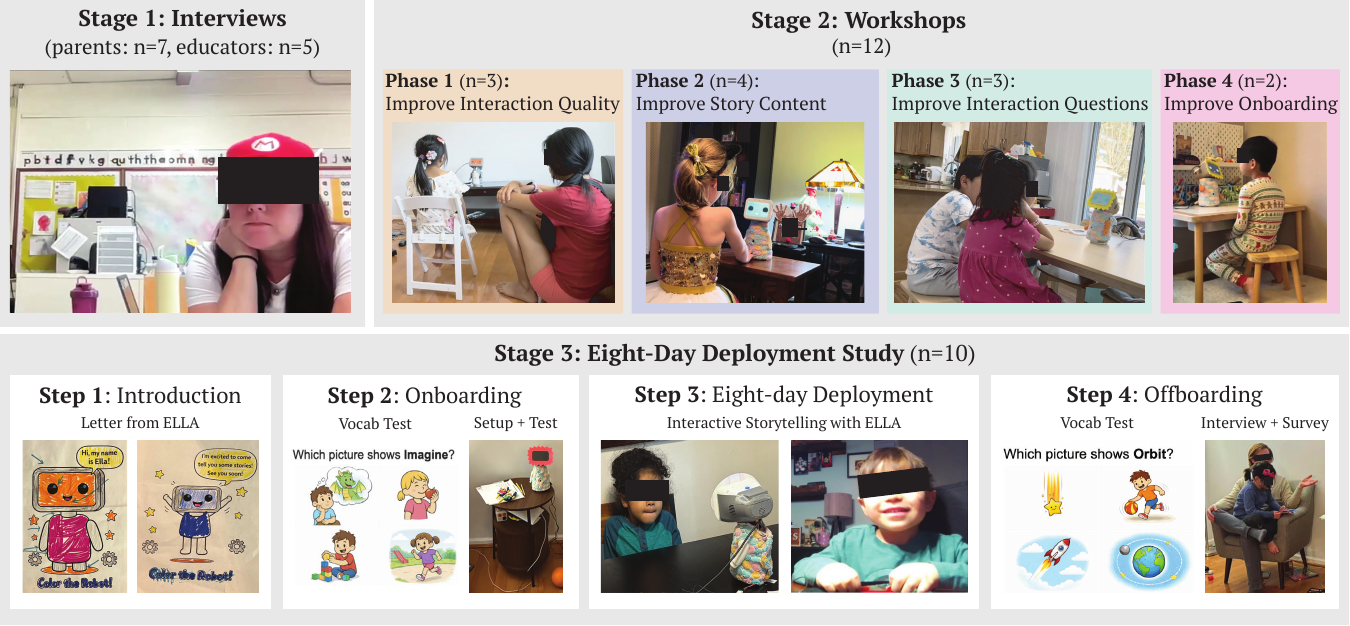}
  \caption{To design a social robot to support language development in children ages 4- 6, we engaged in a three-stage design process involving: 1) online interviews; 2) in-home workshops; and 3) in-home deployment study.}
  \Description{This figure outlines a three-stage design process used to develop a social robot to support language development in children aged 4–6.
  Stage 1 consists of online interviews with parents (n=7) and educators (n=5) to understand needs and inform the initial system design.
  Stage 2 consists of in-home workshops (n=12) organized into multiple iterative phases: Phase 1 (n=3) focuses on improving interaction quality; Phase 2 (n=4) focuses on improving story content; Phase 3 (n=3) focuses on improving interaction questions; and Phase 4 (n=2) focuses on improving onboarding.
  Stage 3 is an eight-day in-home deployment study (n=10) with four steps: (1) an introduction letter from ELLA; (2) onboarding with a vocabulary test and setup; (3) eight days of interactive storytelling with ELLA; and (4) offboarding with a vocabulary test, interview, and survey.}
  \label{fig:overview}
\end{figure}

\subsection{Social Robots for Children's Learning}

Social robots are designed to provide assistance through social rather than physical interaction, offering emotional, cognitive, and social support to promote development, learning, or therapeutic progress \cite{mataric2016socially}. They hold promise in providing support in contexts where human support is constrained by time, attention, or availability. Social robots have demonstrated positive effects on children’s learning, development, and well-being by leveraging their physical embodiments and rich social presences to deliver engaging interventions \cite{belpaeme2018social, scassellati2012robots, robaczewski2021socially, scassellati2018improving,ho2021robomath}. Importantly, social robots can offer consistent and nonjudgmental interactions, which can reduce social pressure while maintaining motivational benefits for children \cite{wright2025robotic}. Their physical embodiment also enables greater engagement and social influence on users in socially assistive tasks than virtual agents \cite{bainbridge2008effect, kidd2005comparison}. Moreover, their physical presence elicits richer social behaviors from children that are beneficial to learning, which translates to enhanced learning outcomes and more sustained engagement compared to virtual agents \cite{leyzberg2012physical, kennedy2015comparing}.   

Significant evidence exists suggesting that social robots, voice assistants, and tablet-based agents can support children’s language learning, especially word learning, story understanding, and story retelling \cite{van2019social}. For example, children ages 4–6 have been shown to learn comparable amounts from a robot, an adult, and a tablet over two sessions, while expressing a strong preference for learning with the robot \cite{westlund2015comparison}. Longer interventions also show promise: compared to a tablet-based program, a robot-based intervention led to greater gains in pre-K children’s storytelling ability, word recognition, and story retelling over two months \cite{hsiao2015irobiq}. For older children (ages 10–12), a two-week in-home study found that a reading-companion robot could support reading comprehension and motivate reading \cite{michaelis2018reading}.

Despite these encouraging signals, social robots for early language learning still face notable limitations. Most systems rely on fixed curricula and highly structured activities, often screen-based or narrow question-and-answer routines, designed around constrained skill targets. This limits the scalability of robot-driven interventions and restricts the robot’s ability to generate expressive, contextually relevant content and to adapt interaction in response to children’s communication and support expressive language use. Yet responsive, conversational interaction is central to creating a rich literacy environment that supports language development \cite{levickis2023associations}. Moreover, much of the empirical work on language-learning robots has been conducted in labs or classrooms \cite{van2019social, kanero2018social}, leaving open questions about how these systems integrate into everyday routines and learning opportunities at home.

In this work, we address these gaps through a three-phase human-centered design process (see Fig. \ref{fig:overview}) featuring stakeholder interviews, in-home workshops, and an in-home deployment using generative AI-powered social robots to enable more flexible, responsive, and ultimately scalable storytelling-based interactions aimed at supporting early language development at home through longer-term child–robot interactions.

\begin{table*}[t]
\centering
\caption{Family Demographics. (*) indicate siblings. ASD stands for autism spectrum disorder. ADHD stands for Attention-deficit/hyperactivity disorder. 50\% of area median income (AMI) is a standard benchmark used by local government programs to define very low-income households, while 80 percent of AMI commonly defines low-income households \cite{HUDPublicHousing, MMP_AMI_Limits, MarylandLegalAidIncomeGuidelines}. Age indicates age at the start of the study.}
\small
\resizebox{\textwidth}{!}{%
\begin{tabular}{p{1.4cm}|p{0.4cm}|p{0.8cm}|p{1.3cm}|p{2.8cm}|p{1.3cm}|p{1.3cm}|p{0.8cm}|p{0.8cm}|p{1.4cm}|p{1.4cm}|p{1.2cm}}

\textbf{Pseudo- nyms} &
\textbf{Age} &
\textbf{Gender} &
\textbf{Ethnicity} &
\textbf{Behavioral \& learning differences} &
\textbf{Household size} &
\textbf{Household income} &
\textbf{50\% AMI} &
\textbf{80\% AMI} &
\textbf{Days/week reading (Eng.)} &
\textbf{Minutes/day reading (Eng.)} &
\textbf{Phases involved} \\
\Xhline{3\arrayrulewidth}
P1-Grace* & 4 & female & Asian & typical & 6 & \$100k--\$125k & above & below & 5 days & 10--20 min & 1, 2, 3 \\
\hline
P2-Andrew & 4 & male & Black and White & speech delay, speech therapy, undergoing diagnosis for behavioral \& neurological differences & 6 & \$50k--\$75k & below & below & 4 days & 10--20 min  & 1, 2, 3 \\
\hline
P3-Sarah & 4 & female & White & typical & 3 & $>$\$200k & above & above & 7 days & 10--20 min  & 1, 2, 3 \\
\hline
P4-Sam & 4 & male & Asian & speech therapy & 4 & \$100k--\$125k & above & above & 2 days & 10--20 min & 1 \\
\hline
P5-Tom & 4 & male & White & typical & 4 & prefer not to answer & above & above & 7 days & $>$30 min & 2 \\
\hline
P6-George & 4 & male & Multiethnic & typical & 3 & \$150k--\$200k & above & above & 7 days & 0--10 min & 3 \\
\hline
P7-Jason & 5 & male & Asian & undergoing diagnosis for ASD \& high intelligence \& ADHD & 4 & \$100k--\$125k & above & below & 7 days & 20--30 min & 1, 2, 3 \\
\hline
P8-Nicole & 5 & female & Asian & typical; English as second language & 4 & $<$\$10k & below & below & 0 days & 0--10 min & 1 \\
\hline
P9-Natalie & 5 & female & Hispanic & typical & 4 & $>$\$200k & above & above & 7 days & 20--30 min & 3 \\
\hline
P10-James* & 6 & male & Asian & used to use ACC device, ASD, speech therapy & 6 & \$100k--\$125k & above & below & 5 days & 0--10 min  & 1, 2, 3 \\
\hline
P11-Susan & 6 & female & Asian & advanced reader & 7 & prefer not to answer & above & above & 7 days & 10--20 min  & 1, 2, 3 \\
\hline
P12-Helen & 6 & female & Multiethnic & ASD, language delay, speech therapy, advanced reader & 4 & \$150k--\$200k & above & above & 7 days & $>$30 min  & 1, 2, 3 \\
\hline
P13-David & 6 & male & Asian & typical & 4 & prefer not to answer & above & above & 7 days & $>$30 min  & 2 \\
\hline
P14-Isabella & 6 & female & Asian & typical & 5 & $>$\$200k & above & above & 7 days & $>$30 min & 2, 3 \\
\end{tabular}
}
\label{tab:children-demographics}
\end{table*}

\section{Understanding the Design Context}

\begin{table}[t]
\centering
\caption{Teacher Demographics. GE stands for General Education. SE stands for Special Education. ESL stands for English as a Second Language.}
\small
\resizebox{\columnwidth}{!}{%
\begin{tabular}{p{0.2cm}|p{0.8cm}|p{1.0cm}|p{1.5cm}|p{1.5cm}|p{2.5cm}|p{1.5cm}|p{2.2cm}}
\textbf{P} &
\textbf{Gender} &
\textbf{Race} &
\textbf{Education} &
\textbf{Experience} &
\textbf{School Type} &
\textbf{Student Age} &
\textbf{Student Socio-Economic Status} \\
\Xhline{3\arrayrulewidth}

1 & F & Hispanic & Master’s & 7 yrs &
Public (GE, SE, ESL) & 4--6 & Mixed \\
\hline

2 & F & White & Bachelor’s & 20 yrs &
Public (GE, SE, ESL) & 3--5 & Mixed \\
\hline

3 & F & White & Master’s & 40 yrs &
Private (GE) & 4--11 & High \\
\hline

4 & F & Asian & Bachelor’s & 3.5 yrs &
Private (GE, ESL) &
3--4; 14--18 & High \\
\hline

5 & M & Black & Master’s & 7 yrs &
Public Charter (GE, SE, ESL) &
7--8; 10--14 & Mixed \\

\end{tabular}
}
\label{tab:interview-demographics-teacher}
\end{table}

To ground our design process in the lived experiences of stakeholders involved in early language development, we conducted semi-structured interviews with parents ($n=7$, see demographics in Table \ref{tab:children-demographics}) and educators ($n=5$, see demographics in Table \ref{tab:interview-demographics-teacher}). These interviews focused on participants’ everyday language-learning practices, observed challenges, instructional strategies, and the constraints shaping language use at home and in classrooms. Parent interviews emphasized family routines, reading practices, and strategies used to support children’s language development in daily life, while educator interviews focused on classroom-based language instruction, vocabulary teaching practices, and home–school connections. 

All interview recordings were transcribed and analyzed using a reflective thematic analysis approach. We synthesized themes from these interviews, along with insights from prior work in early language development and human–robot interaction, to derive a set of design goals (DG) to guide the development of a generative AI–powered social robot for at-home early language development.

\textbf{DG1:} \colorbox{lightpurple}{Leverage storytelling as a flexible, interest-driven learning context.}

\textit{Rationale.} Parents and educators consistently described storytelling as a central pedagogical practice for supporting early language development. Stories were selected based on children’s current interests (\eg animals, fantasy) and children were encouraged to retell, elaborate on, or invent their own narratives. These practices align with prior work showing that storybook reading and oral storytelling support vocabulary growth, narrative comprehension, and expressive language, particularly when children are actively engaged \cite{hargrave2000book}. Storytelling represents a dynamic and flexible learning context that leverages children’s interests to support rich language use.

\textbf{DG2:} \colorbox{lightpurple}{Provide repeated, contextualized exposure to vocabulary.}

\textit{Rationale.} Educators described revisiting the same stories or themes across days, while parents reported rereading favorite books, following book series, and repeating words during everyday conversations. This mirrors extensive evidence that effective vocabulary acquisition requires repeated exposures to target words embedded in meaningful discourse and contexts \cite{justice2005learning, pinkham2011you, harris2011lessons}. These practices highlight the importance of prioritizing exposure returning to a set of target words across stories and time to support learning and engagement.

\textbf{DG3:} \colorbox{lightpurple}{Blend explicit and implicit vocabulary instruction.}

\textit{Rationale.} Educators noted introducing and discussing target words during read-alouds, while parents similarly interjected short explanations (\eg “this means…”) without breaking the narrative flow. This approach is well supported in the literature, which shows that blending implicit exposure (\eg modeling use) with explicit definitions (\eg child-friendly definitions) leads to stronger vocabulary learning \cite{coyne2007vocabulary}. These findings emphasize the value of integrating explanation, exposure, and modeling directly into storytelling.

\textbf{DG4:} \colorbox{lightpurple}{Support both receptive and expressive language development.}

\textit{Rationale.} Children often understand more words than they can articulate. To encourage expressive language, educators and parents described using low-pressure prompting strategies (\eg fill-in-the-blank questions) without requiring correctness. Prior developmental research similarly shows that expressive language growth is supported when children are given low-pressure opportunities to speak without evaluative pressure \cite{benedict1979early}. This evidence motivates interaction designs that invite verbal participation, along with allowing children to listen to the narrative.


\textbf{DG5:} \colorbox{lightpurple}{Scaffold language use to support diverse developmental needs.}

\textit{Rationale.} Substantial variability in children’s language abilities exists and is shaped by factors such as bilingual backgrounds, speech therapy, and developmental differences \cite{wang2014exploring}. Educators emphasized the importance of differentiation (\ie adjusting questions, modifying activity) while parents described informally adapting their own prompts based on daily fluctuations in attention and interest. These practices align with Vygotskian notions of scaffolding and the Zone of Proximal Development, which emphasize tailoring support to a learner’s current capabilities \cite{irshad2021vygotsky} and foregrounds the need for interaction designs that can flexibly adjust difficulty and scaffold learning.


\textbf{DG6:} \colorbox{lightpurple}{Maintain engagement through expressive storytelling practices.}

\textit{Rationale.} Engagement emerged as a prerequisite for learning across interviews. Parents and teachers described using expressive prosody, voice modulation, facial expression, and theatrical delivery to maintain children’s attention during storytelling. Prior work in child–robot interaction similarly demonstrates that children respond more positively to robots that exhibit social presence and expressive behavior \cite{kory2017flat, kory2014storytelling}. These findings position expressiveness not as an aesthetic feature, but as a functional requirement for sustaining attention and participation over repeated, at-home interactions.




\section{Iterative Design at Home, with Families}

To shape our early interaction concepts into a concrete, deployable system with interaction design grounded in the lived environment, we engaged in a series of at-home design workshops with families and children. We conducted a series of at-home design workshops with children and their families. These workshops allowed us to iteratively refine an exploratory social robot prototype by observing children’s interactions with early versions of the system and incorporating feedback from both children and parents. Across 12 workshops, we progressively evolved the robot’s interaction mechanics, story structure, and pedagogical content to better support engagement and early language learning at home.

\textbf{Exploratory Prototype.} We developed an exploratory social robot prototype designed to support early language development through interactive storytelling (\textit{DG1}). The robot generated stories using a multi-stage generative AI pipeline and combined spoken narration with expressive non-verbal behaviors (\textit{DG6}). Modeling off of teacher-driven story-based vocabulary interventions (see Sec.~\ref{sec:related-work-early-language-development}), each story (roughly 350 words) was structured around three target vocabulary words. The robot explicitly defined the new words at the beginning of the story and implicitly reinforced all target words throughout the narrative (\textit{DG2,3}). Stories included three interaction points, at which the robot asked questions intended to support vocabulary understanding and use (\textit{DG4}), with three scaffolded follow-up prompts (\textit{DG5}). Each story session lasted about 12--18 minutes. This exploratory prototype served as a technology probe, allowing us to surface design challenges and opportunities through direct use in children’s homes.

\textbf{Workshop Protocol.} After obtaining informed consent from parents and assent from children, we introduced the robot and explained its purpose in age-appropriate terms. Children then experienced one robot-led storytelling session and were given the option to continue with additional stories (up to three total). Following the interaction, we conducted a semi-structured interview with parents and children focused on their experience, engagement, and perceived value. All sessions were video-recorded, and researchers took observational notes to capture interaction breakdowns, moments of engagement, and parent mediation strategies. The study was approved by our IRB. Each workshop took approximately 60 minutes. Families were compensated with a $\$20$ gift card, and children were compensated with robot stickers. 

\textbf{Phase 1: \colorbox{lightorange}{Improving Interaction Flow \& Quality}}

\textit{Observed Breakdowns.} The first three workshops revealed fundamental shortcomings in the robot’s interaction mechanics, dialogue flow, and perceived approachability. Although children were initially curious about the robot, engagement often deteriorated due to repeated interaction breakdowns. High response latency (approximately 10 seconds between children’s speech and the robot’s reply) disrupted conversational turn-taking, leading to interruptions, missed turns, and visible frustration and loss of interest. Children frequently spoke over the robot or disengaged while waiting for responses.

In addition to timing issues, we observed challenges related to the robot’s social presence and physical appearance. Several children appeared hesitant or uneasy around the robot, describing it as ``scary'' or expressing dislike for its appearance, even when they were familiar with voice-based agents such as Alexa. Rather than responding directly to the robot’s questions, children often redirected their answers to a caregiver or researcher, suggesting limited social attribution or comfort with the robot as an interaction partner. The robot’s non-verbal behaviors were perceived as limited in expressiveness and repetitive, further weakening its perceived social presence.

\textit{Design Evolution.} To address conversational breakdowns, we restructured turn-taking from a voice activity detection (VAD) pipeline and LLM-based turn-end detection approach to a VAD and semantic intent model method, enabling faster and more reliable detection of conversational turns. We reduced perceived latency by implementing incremental text-to-speech generation (chunking), allowing the robot to begin responding more quickly. 

To address issues of approachability and social comfort, we redesigned the robot’s physical presentation and onboarding experience. We introduced brighter colors, interchangeable faceplates, and a soft, plush exterior to make the robot appear less intimidating. We also added a short onboarding interaction in which children were invited to touch the robot, select and attach a faceplate, and add simple accessories. This participatory introduction helped children acclimate to the robot, increased willingness to engage directly, and shifted responses from caregivers toward the robot itself.

\textbf{Phase 2: \colorbox{lightblue}{Improving Story Content \& Flow}}

\textit{Observed Breakdowns.} Once interaction timing improved, workshops 4–6 surfaced challenges with story structure, complexity, and cognitive load. Children struggled to follow longer, conceptually dense stories that introduced multiple vocabulary words. For example, early stories situated learning within imaginative settings (\eg space) and introduced several target concepts (\eg moon, crater, gravity) within a single narrative. While initially engaging, these stories proved difficult for children of this age to comprehend, limiting attention and comprehension.

\textit{Design Evolution.} We simplified story design by shortening narratives, reducing conceptual complexity, and limiting each story to a single target vocabulary word. Stories were reframed around familiar, everyday contexts (\eg playing on the playground) and relatable experiences personalized to the child’s thematic interests (\eg dolphin); these changes aimed to reduce cognitive load, making it easier for children to follow the plot and attend to the meaning of the target word. We refined the text-to-speech model (GPT-4o mini TTS) prompt to slow down the robot's speech and enrich the story's delivery. In parallel, we expanded the robot’s expressive behavior set and refined its behavior prompts to improve the smoothness, variation, and timing of gestures and affective cues during storytelling, with a focus on teaching the target word.

\textbf{Phase 3: \colorbox{lightcyan}{Improving Interaction Questions}}

\textit{Observed Breakdowns.} In early versions, interaction questions were interspersed throughout the story and often required abstract reasoning (\eg imagining what would happen without gravity). These questions disrupted narrative flow and confused children, particularly before they had fully understood the story. Workshops 7-9 focused on shaping and sequencing interactions to establish story comprehension before prompting vocabulary use.

\textit{Design Evolution.} We restructured interaction points to align with children’s comprehension processes. Questions were repositioned to follow the narrative arc, progressing from story perception (e.g., enjoyment), to story recall, and finally to vocabulary practice grounded in personal experience (e.g., “Have you ever been in an apartment?”). This sequencing supported comprehension before prompting vocabulary use. We also refined the robot scaffolding strategies and prompts to match the three types of questions.

\textbf{Phase 4: \colorbox{lightpink}{Improving Onboarding and Deployability}}

\textit{Design Evolution.} In Phases $1$--$3$, researchers set up the system and loaded the stories manually using a terminal attached to the robot. In the final workshops $10$--$12$, we focused on stabilizing the system for home deployment. We developed a user interface for ease of generating and reviewing individualized interaction content (stories and interaction questions) for deployment. We added content moderation using Llama-Guard-4-12B for detecting unsafe content from real-time LLM-generated responses. When unsafe content is detected, the interaction handler LLM will be prompted to regenerate a new response. We refined onboarding procedures (\ie how to introduce ELLA to children), improved the end-to-end session flow (added wake-up and sleep animations and automatic shutdown), and validated data collection and logging pipelines. These deployment-oriented refinements informed the final system used in our in-home deployment study.

\section{\textit{ELLA}: Our Deployment-Ready Prototype}

\textbf{Social Robot Platform}
ELLA is built on a custom social robot platform actuated by five servo motors controlling base rotation, head tilt and rotation, and independent arm movement. All components are housed in a 3D-printed shell. 



\textbf{Interaction Flow} A physical on/off button on the robot initiates the interactive storytelling session. While the robot boots, ELLA plays a short wake-up animation; in parallel, it fetches that day’s four stories from the cloud story database. Once ready, ELLA greets the child and asks whether they would like to hear a story. Each story is generated around a parent-selected target word and a theme aligned with the child’s interests. ELLA then delivers the story and its embedded questions, and asks if the child would like to hear another. This loop continues until the child declines or ELLA has delivered all four stories scheduled for the day. When the session ends, ELLA plays a sleep animation while background processes shut down. If ELLA is turned on again later and stories remain for the day, it resumes with the next story. If no stories remain, ELLA tells the child it has run out of stories and is “dreaming up” new ones before returning to sleep.

\begin{figure}[t]
  \includegraphics[width=\textwidth]{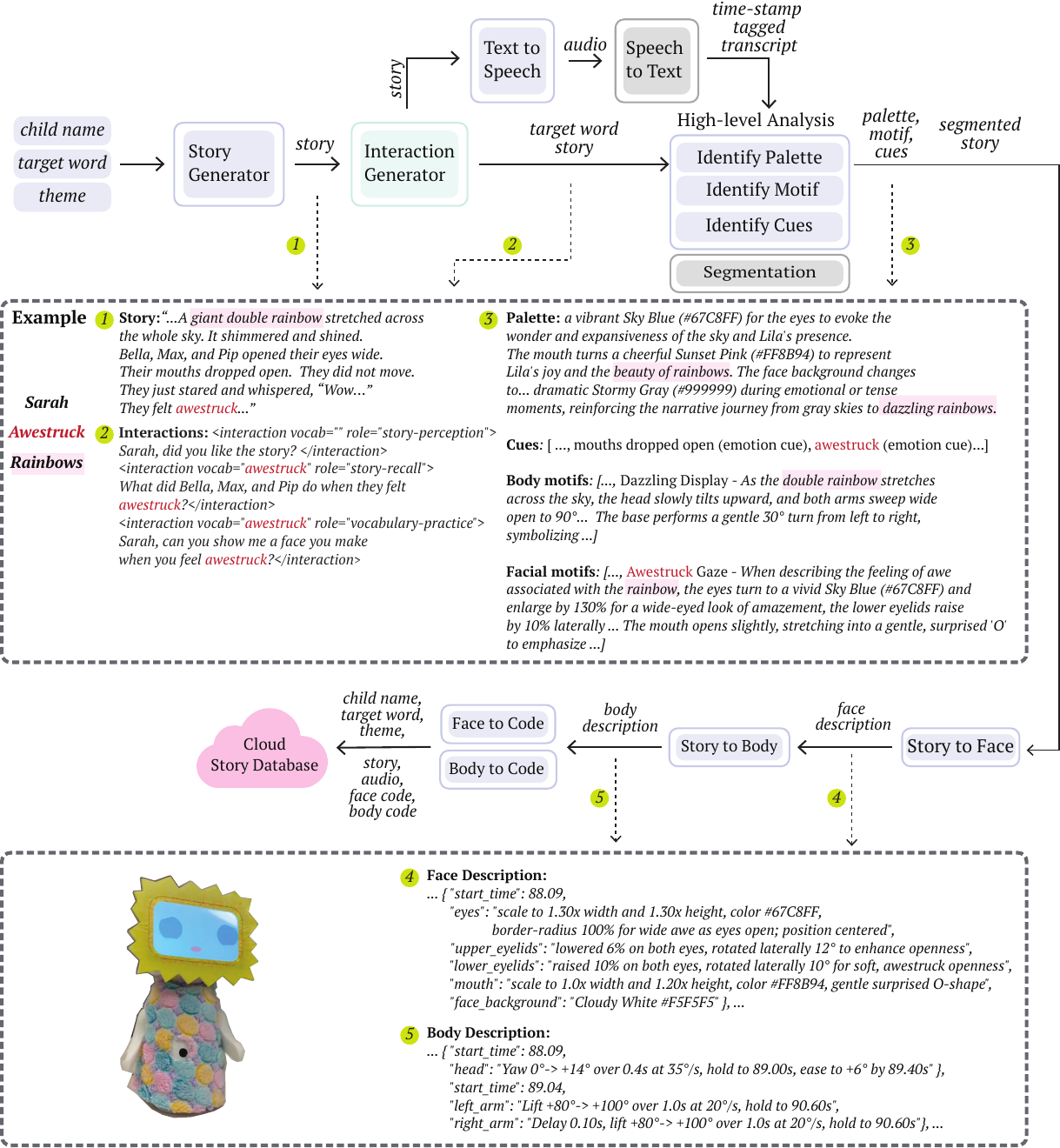}
  \caption{Pipeline for generating personalized and targeted content for interactive storytelling with ELLA. Design evolution (\colorbox{lightorange}{Phase 1}, \colorbox{lightblue}{Phase 2}, \colorbox{lightcyan}{Phase 3}, \colorbox{lightpink}{Phase 4}) of components indicated by color.}
  \Description{This figure presents the ELLA Interaction Content Generation Pipeline, with color-coded components indicating design evolution across Phases 1–4.
  The pipeline begins with inputs including the child’s name, target vocabulary word, and story theme, which are used to generate a story and associated interactions. The pipeline then branches, with one path converting the story to audio and producing a timestamped transcript, and the other passing the story and target word directly forward.
  Both paths feed into a high-level analysis stage that identifies interaction elements such as palettes, motifs, cues, and story segments. These elements are used to generate facial and bodily behaviors, which are converted into executable code and stored, along with the story and audio, in a cloud-based story database.
  Each stage in the diagram includes example artifacts illustrating the information passed between components.}
  \label{fig:generation-pipeline}
\end{figure}

\begin{figure}[t]
  \includegraphics[width=\textwidth]{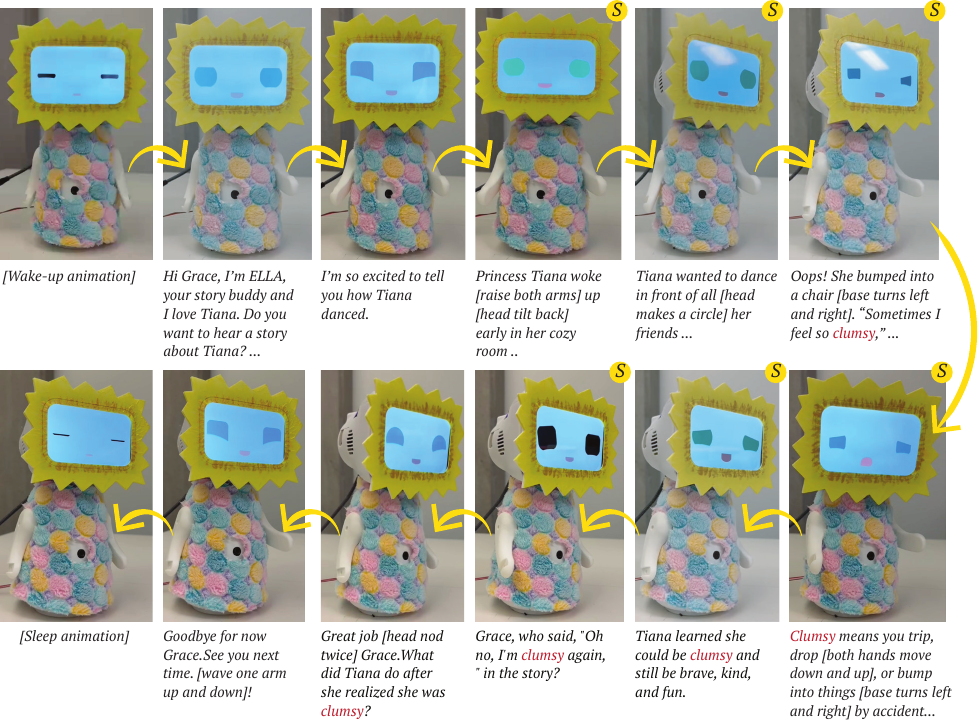}
  \caption{Example story and behaviors. S denotes behaviors generated with the storytelling behavior generation pipeline. The rest of the behaviors are generated using the interaction behavior generation pipeline. Child name is replaced with pseudonym.}
  \Description{This figure shows an example storytelling session with ELLA, including both spoken content and the robot’s accompanying behaviors. The child’s name is replaced with a pseudonym.
  The session begins with a greeting and proceeds through storytelling, during which the robot performs synchronized gestures (e.g., arm, head, and base movements) and changes facial expressions to emphasize the narrative and target vocabulary. The interaction also includes comprehension questions and feedback, and concludes with a farewell and sleep animation.
  Behaviors marked with S are generated by the storytelling behavior generation pipeline, while the remaining behaviors are generated by the interaction behavior generation pipeline. Each interaction state is paired with a corresponding image of the robot.}
  \label{fig:example-story}
\end{figure}

\subsection{Story Generation}
The content for each ELLA storytelling session is generated through a two-step, language model-driven pipeline (see Fig. \ref{fig:generation-pipeline}): story generation and interaction authoring.

\textbf{Step 1: Story Authoring.} Given a theme and a target vocabulary word, a GPT-5 language model is prompted to write a 200-word story for children ages 4--6 following a fixed narrative arc (exposition → conflict → resolution). The prompt constrains the story so that the theme is central to the setting and plot, the target word appears at least three times, and its meaning is taught both implicitly (through story events) and explicitly (via a child-friendly in-story definition). The prompt also constrains style (short sentences, simple words, warm tone) and prohibits adding the child as a character.

\textbf{Step 2: Interaction Authoring.} A second GPT-5 language model takes the generated story text as input and produces a script that repeats the story verbatim, followed by exactly three questions in a fixed pedagogical sequence: (1) a low-pressure story-perception question (e.g., whether the child liked the story), (2) a story-recall question that references a specific event and naturally incorporates the target word, and (3) a vocabulary-practice question that prompts the child to use the target word in a new, relatable context. During the session, the story-recall and vocabulary-practice questions each trigger two follow-up questions generated on the fly.


\subsection{Storytelling Behavior Generation}

ELLA delivers each story with expressive, context-appropriate multimodal non-verbal behaviors aligned with its verbal narration (see Fig. \ref{fig:example-story} for example). We pre-generate these expressive behaviors with our \textit{Xpress3D} pipeline (see Fig. \ref{fig:generation-pipeline}).

\textbf{Xpress3D.} We extend Xpress \cite{antony2025xpress} (a language-model-driven pipeline for context-aware robot facial expressions generation)  into a multimodal behavior generator that produces expressive robot non-verbal behaviors (\ie facial expression and body gestures) aligned to a story’s narrative structure, tone, and context.

We start from a word-level, timestamped transcript of the story audio; a GPT-5 language model (LM) segments this transcript into narratively meaningful chunks and generates a segment-level color palette that serves as meta-context. We add an expressiveness planning layer conditioned on this meta-context: for each segment, we (i) identify lexical/narrative cues that represent gesture opportunities (e.g., emotion, action, emphasis, spatial language), and (ii) use these cues, together with segment context and the palette, to generate high-level, timestamped behavior descriptions. Each segment’s behavior is generated sequentially, conditioned on prior segments and previously generated behaviors, which supports narrative coherence and smooth evolution across the story.

Lastly, we compile high-level descriptions into executable robot programs for each channel. A GPT-5 model translates the face descriptions into executable anime.js programs for the robot's animated face, using prompt-specified element capabilities plus explicit rules and stepwise verification to reduce errors and avoid abrupt transitions. For the body, we introduce an analogous code-generation prompt that outputs a time-indexed trajectory of joint “stages” controlling five robot joints (head yaw/pitch, two arms, base rotation) with hard limits and minimum-duration constraints to maintain safety and smoothness.

\begin{figure}[t]
  \includegraphics[width=\textwidth]{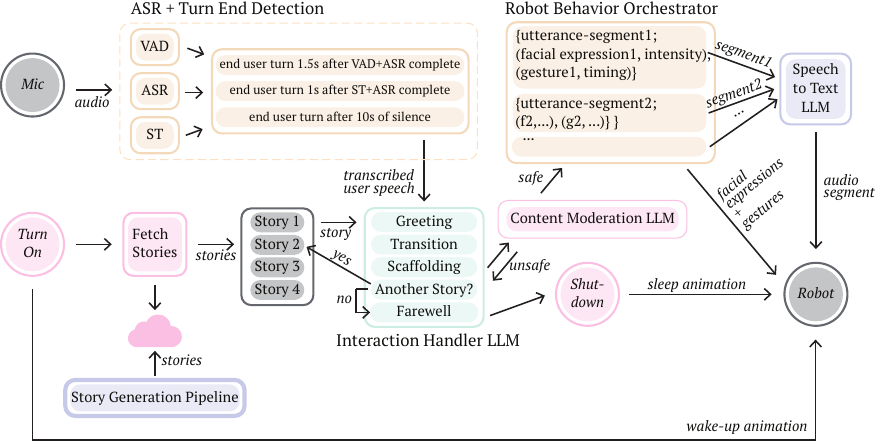}
  \caption{ELLA System Diagram. Design evolution (\colorbox{lightorange}{Phase 1}, \colorbox{lightblue}{Phase 2}, \colorbox{lightcyan}{Phase 3}, \colorbox{lightpink}{Phase 4}) of components indicated by color.}
  \Description{This figure presents a system diagram of ELLA, illustrating how speech input, story content, and interaction logic are integrated to generate robot behaviors. Color-coded components indicate design evolution across Phases 1–4.
  Considered as a whole, the system takes audio input from the child, processes it to determine turn-taking and interaction state, and uses language models to manage storytelling flow, moderation, and dialogue decisions. Story content is retrieved or generated and combined with high-level interaction handling to determine whether to continue storytelling, scaffold responses, or conclude the session.
  Based on these decisions, the system generates coordinated facial expressions, gestures, and animations, which are executed by the robot. The diagram highlights the end-to-end flow from user speech and story content to robot behaviors, emphasizing modular design and interaction safety.}
  \label{fig:system}
\end{figure}

\subsection{Interaction Behavior Generation}
ELLA uses an LLM-based approach to drive real-time interactions and to deliver responses with expressive, context-appropriate non-verbal behaviors picked from a fixed behavior library, enabling real-time multimodal responses (see Fig. \ref{fig:system}).

When an interaction chunk begins, the robot delivers the chunk’s first question together with its pre-generated behaviors, then hands the floor to the child. The child’s speech is captured by continuously buffering partial ASR into a running transcript while a lightweight turn manager detects end-of-turn using both voice-activity detection and semantic turn-end signals. Specifically, ELLA waits up to 10 seconds for speech onset; once speech begins, it ends the turn after 1.5 seconds of silence, or—if a turn-end detector reports \textit{complete} and ASR text is present—after an additional 0.5 seconds grace period to capture trailing speech. The finalized child utterance is then committed to conversation memory and used for response generation.

For each child turn, a GPT-OSS 120B LLM produces a structured behavior plan consisting of a selected scaffolding/teaching strategy followed by one or more short utterance items. Each item specifies (i) the text to speak, (ii) a facial expression with intensity, and (iii) a gesture primitive with coarse timing. To keep latency low, ELLA chunks text-to-speech requests and begins playback as soon as the first audio chunk is available, while executing the planned facial expression and gesture in sync with the utterance. Constraining the LLM to a banked set of facial expressions and gesture primitives makes planning reliable and fast, while still providing a sufficiently diverse expressive repertoire.

\section{In-Home Deployment}
To examine how children interact with a storytelling robot in real-world settings, how the system fits into family routines, and whether it can support early language development, we conducted an in-home deployment of ELLA with ten children (see Table \ref{tab:children-demographics}). Each family hosted the robot for eight consecutive days, allowing us to observe use over repeated interactions and within everyday household contexts. Comparable to prior research-based vocabulary interventions for this age group, ELLA is designed to teach $4$ target words over 8 days. 

\subsection{Procedure}
Prior to deployment, parents selected four target vocabulary words that their child did not yet know but that they believed would be beneficial to learn. Parents also provided story themes based on the child’s interests (e.g., favorite books, movies, television shows, activities, or topics). To build anticipation, families were mailed a letter written from ELLA’s perspective along with a “color-me” activity featuring the robot \cite{lee2022unboxing}. We generated one story using one target word and a randomly selected theme for each day. To ensure the safety of the generated content, the research team manually reviewed all the stories and pre-generated interaction questions for safety and child appropriateness prior to deployment. 

ELLA was delivered to each home by two researchers, who set up the robot and provided an in-person orientation. Families also received a printed user manual detailing basic operation (e.g., powering the robot on and off) and data practices (e.g., when audio and video recording occurred). Parents were encouraged to engage with ELLA for at least one story per day but were explicitly told that interaction timing and frequency were flexible, and that missing days was acceptable. This framing was intended to reduce pressure and allow families to integrate the robot naturally into their routines.

Throughout the deployment, parents were asked to complete a semi-structured daily diary documenting their observations. To support compliance, parents received a reminder midway through the deployment (day four) to continue completing diary entries. 

After eight days, researchers returned to retrieve the robot and conduct off-boarding assessments and interviews. To support closure and robot handoff, children and their siblings received a printed photo of themselves with ELLA and a small 3D-printed ELLA toy. 


\subsection{Data Sources} 

We leveraged multiple data sources to capture learning outcomes, interaction patterns, and family experiences.

\textbf{Vocabulary Assessment.} We conducted a pre- and post-deployment receptive vocabulary assessment using a Peabody Picture Vocabulary Test (PPVT)–style\cite{dunn1965peabody} format to determine whether children knew their selected target words at baseline and after the deployment. For each target word, we generated a custom PPVT-style item using GPT-5 (see Fig. \ref{fig:overview} for an example).

\textbf{Parent Diaries.} Parents completed a short daily diary throughout the eight-day deployment, including on days when the child did not use ELLA. The diary captured what initiated sessions (child-initiated, parent-prompted, routine-based), who was present, and the child’s affect. It also asked about spillover beyond sessions and whether the child used any target words outside the interaction. On non-use days, parents indicated reasons (e.g., time constraints, lack of interest, fatigue, technical issues), with space for open-ended notes.

\textbf{Post-Deployment Interviews.} After the deployment, we conducted semi-structured interviews with parents and structured interviews with children. Parent interviews focused on overall experience, changes to family routines, perceived learning, breakdowns, and desired improvements. Child interviews used an emoji-based response scale and open prompts to capture children’s affective responses, perceived understanding, enjoyment, and recall of stories. Because most children were pre-literate, all questions and response options were read aloud by a researcher (see interview instruments in the supplemental materials \footnote{link to supplementary materials: \url{https://intuitivecomputing.github.io/publications/2026-antony-ella-supp.pdf}}).

\textbf{Robot Session Logs and Recordings.} ELLA recorded audio and video during story sessions, logged session start and end times, and stored system logs with conversational transcripts. These logs supported analysis of interaction duration, frequency, and dialogue structure.

\begin{figure}[t]
  \includegraphics[width=\textwidth]{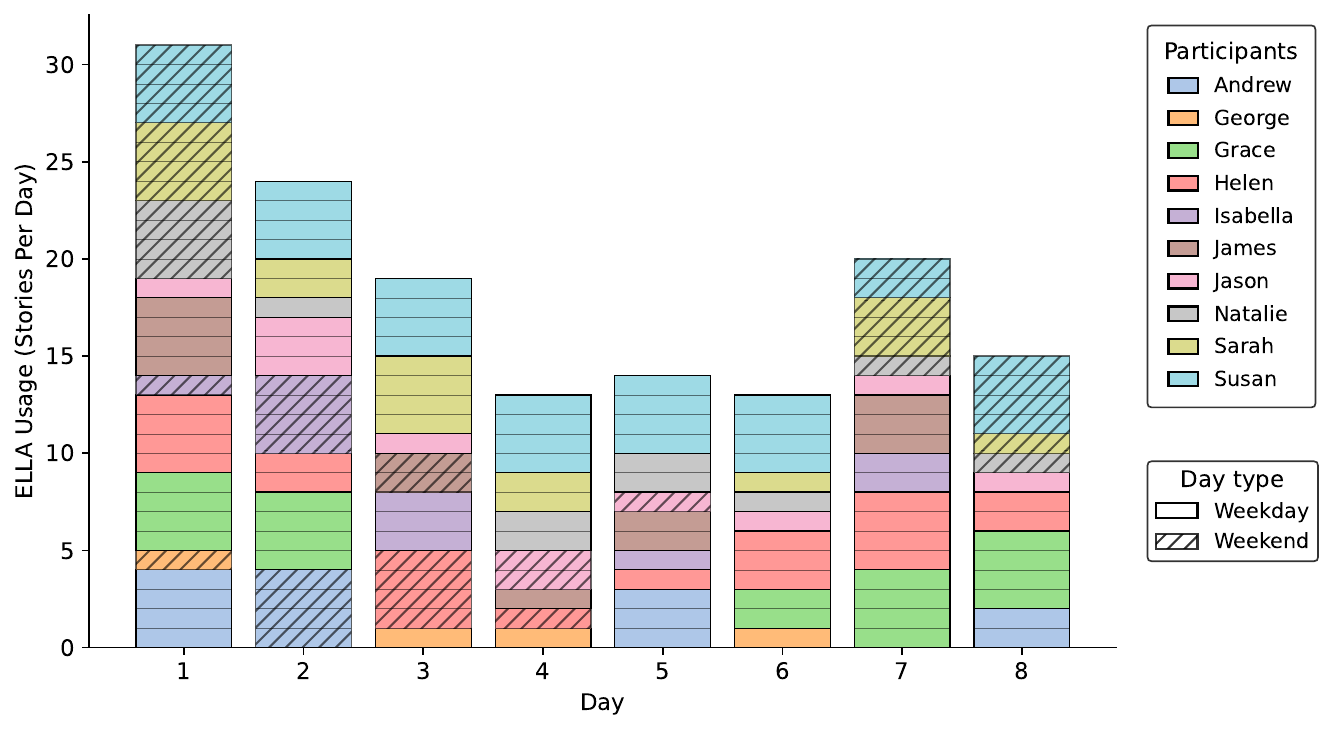}
  \caption{ELLA Usage over time.}
  \Description{This figure shows a bar chart of ELLA usage over time, measured as the number of stories told per day for each child across an eight-day deployment period. The horizontal axis represents the day of deployment, and the vertical axis represents the number of stories per day.
  Each color of bars corresponds to an individual child, illustrating day-to-day variation in usage. Bars with diagonal hatching indicate weekend days, during which usage patterns differ from weekdays. Across children, usage ranges from zero to four stories per day, reflecting differences in engagement frequency over time.
  The chart highlights both periods of consistent daily use and days with no usage, providing an overview of how storytelling activity varied across participants and across the deployment duration.}
  \label{fig:ella-usage}
\end{figure}

\begin{table*}[t]
\centering
\caption{Context based on parent diaries.}
\small
\begin{tabular}{p{2.4cm} r p{2.4cm} r p{5.5cm} r}
\textbf{Who Initiated} & \textbf{n} &
\textbf{Who Were Present} & \textbf{n} &
\textbf{How Child Felt About Storytelling Session} & \textbf{n} \\
\hline
Parent prompted & 34 & Parent & 42 & Neutral & 23 \\
Scheduled routine & 19 & Sibling & 27 & Excited & 20 \\
Child initiated & 16 & Child alone & 5 & Frustrated & 7 \\
Sibling encouraged & 7 & Other caregivers & 2 & Hesitant & 6 \\
Friend initiated & 1 & Friend & 1 & Want to continue beyond session & 3 \\
 &  &  &  & Happy & 1 \\
\end{tabular}
\label{tab:initiation-presence-feeling}
\end{table*}

\begin{figure}[t]
  \includegraphics[width=\textwidth]{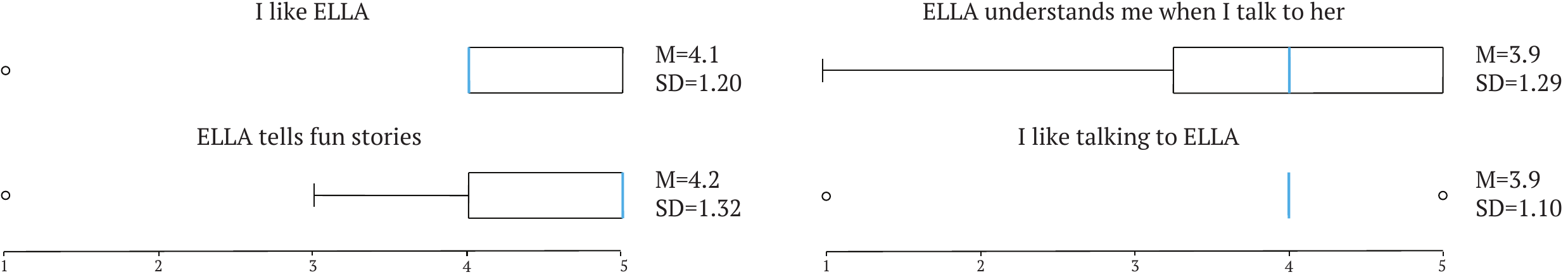}
  \caption{Child survey results. The blue line marks the median.}
  \Description{This figure presents results from a child survey evaluating interactions with ELLA. The survey includes items assessing enjoyment and perceived understanding, such as liking ELLA, finding the stories fun, feeling understood when talking to ELLA, and enjoying talking to ELLA.
  For each item, responses are summarized using descriptive statistics, with mean scores ranging from 3.9 to 4.2 and standard deviations between 1.1 and 1.32. The blue line indicates the median response for each survey item, providing a visual reference for central tendency across participants.}
  \label{fig:survey-results}
\end{figure}

\begin{table*}[t]
\centering
\caption{Target words and themes for each child during deployment. Green highlight represents words that the child learned by the end of the deployment study.}
\small
\resizebox{\textwidth}{!}{%
\begin{tabular}{p{1.4cm}|p{0.6cm}|p{1.5cm}|p{1.5cm}|p{1.5cm}|p{1.5cm}|p{3.1cm}|p{1.2cm}|p{1.2cm}}
\textbf{Pseudonym} &
\textbf{Age} &
\multicolumn{4}{c|}{\textbf{Target Words}} &
\textbf{Example Theme} &
\textbf{Days used ELLA} &
\textbf{Stories listened to} \\
\Xhline{3\arrayrulewidth}
Grace & 4 &
Imitate & Ordinary &
\cellcolor[HTML]{D0EBAD}Clumsy &
\cellcolor[HTML]{D0EBAD}Massive &
Rapunzel (disney princess) & 5 & 18 \\

\hline
Andrew & 4 &
\cellcolor[HTML]{D0EBAD}Imagine &
Confident &
\cellcolor[HTML]{D0EBAD}Permission &
Self-control &
Tom and Jerry (TV show) & 4 & 13 \\

\hline
Sarah & 4 &
\cellcolor[HTML]{D0EBAD}Perseverance &
\cellcolor[HTML]{D0EBAD}Compassion &
\cellcolor[HTML]{D0EBAD}Gumption &
\cellcolor[HTML]{D0EBAD}Awestruck &
Pony & 7 & 18 \\

\hline
George & 4 &
Permission &
Consequences &
Orbit &
\cellcolor[HTML]{D0EBAD}Chirp &
Ironman & 2 & 4 \\

\hline
Jason & 5 &
\cellcolor[HTML]{D0EBAD}Imitate &
\cellcolor[HTML]{D0EBAD}Somersault &
\cellcolor[HTML]{D0EBAD}Clumsy &
\cellcolor[HTML]{D0EBAD}Frisky &
Science & 8 & 11 \\

\hline
Natalie & 5 &
Apartment &
Advocate &
Justice &
\cellcolor[HTML]{D0EBAD}Bait &
Dance & 7 & 12 \\

\hline
James & 6 &
\cellcolor[HTML]{D0EBAD}Permission &
Sympathy &
\cellcolor[HTML]{D0EBAD}Wonder &
\cellcolor[HTML]{D0EBAD}Frisky &
Truck & 4 & 10 \\

\hline
Susan & 6 &
\cellcolor[HTML]{D0EBAD}Considerate &
Persistent &
\cellcolor[HTML]{D0EBAD}Attempt &
\cellcolor[HTML]{D0EBAD}Achieve &
Bunny & 8 & 30 \\

\hline
Helen & 6 &
\cellcolor[HTML]{D0EBAD}Advocate &
\cellcolor[HTML]{D0EBAD}Legal &
\cellcolor[HTML]{D0EBAD}Empathy &
\cellcolor[HTML]{D0EBAD}Sympathy &
Polar Express (movie) & 8 & 20 \\

\hline
Isabella & 6 &
\cellcolor[HTML]{D0EBAD}Adventure &
\cellcolor[HTML]{D0EBAD}Orbit &
\cellcolor[HTML]{D0EBAD}Usual &
\cellcolor[HTML]{D0EBAD}Sheriff &
Minecraft & 5 & 11 \\

\end{tabular}
}
\label{tab:target-words-themes}
\end{table*}

\begin{table*}[t]
\centering
\caption{Child Language usage during deployment. Green highlight represents words that the child learned by the end of the deployment study.}
\small
\resizebox{\textwidth}{!}{%
\begin{tabular}{p{1.4cm}|p{0.6cm}|p{1.0cm}|p{1.0cm}|p{8.0cm}|p{1.0cm}|p{1.4cm}}
\textbf{Pseudonym} &
\textbf{Age} &
\textbf{Stories listened to} &
\textbf{Target words used} &
\textbf{Target word usage during story sessions} &
\textbf{Total target word used} &
\textbf{Avg words/ turn} \\
\Xhline{3\arrayrulewidth}
Grace     & 4 & 18 & 3/4 & \colorbox{green}{Massive} (14), Ordinary (16), \colorbox{green}{Clumsy} (13), Imitate (0) & 43 & 3.78 \\
\hline
Andrew    & 4 & 13 & 3/4 & \colorbox{green}{Permission} (3), Self-Control (2), \colorbox{green}{Imagine} (1), Confident (0) & 6 & 3.48  \\
\hline
Sarah    & 4 & 18 & 4/4 & \colorbox{green}{Compassion} (2), \colorbox{green}{Awestruck} (2), \colorbox{green}{Perseverance} (0), \colorbox{green}{Gumption} (0) & 4 & 3.48  \\
\hline
George    & 4 & 4  & 1/4 & \colorbox{green}{Chirp} (2), Permission (0), Consequences (0), Orbit (0) & 2 & 2.46\\
\hline
Jason     & 5 & 11 & 2/4 & \colorbox{green}{Clumsy} (8), \colorbox{green}{Imitate} (1), Somersault (0), Frisky (0) & 9 & 4.91\\
\hline
Natalie  & 5 & 12 & 2/4 & Advocate (2), \colorbox{green}{Bait} (1), Justice (1), Apartment (0) & 4 & 3.02 \\
\hline
James     & 6 & 10 & 3/4 & \colorbox{green}{Frisky} (4), \colorbox{green}{Wonder} (3), \colorbox{green}{Permission} (1), Sympathy (0)& 8 & 2.48  \\
\hline
Susan     & 6 & 30 & 4/4 & \colorbox{green}{Achieve} (23), \colorbox{green}{Attempt} (17), Persistent (16), \colorbox{green}{Considerate} (10) & 66 & 6.30  \\
\hline
Helen     & 6 & 20 & 4/4 & \colorbox{green}{Advocate} (9), \colorbox{green}{Sympathy} (7), \colorbox{green}{Legal} (5), \colorbox{green}{Empathy} (5) & 26 & 4.98 \\
\hline
Isabella  & 6 & 11 & 3/4 & \colorbox{green}{Usual} (3), \colorbox{green}{Sheriff} (1), \colorbox{green}{Adventure} (1), Orbit (0) & 5 & 4.60  \\
\end{tabular}
}
\label{tab:target-word-usage}
\end{table*}

\subsection{Results}
Across the eight-day deployment, children used ELLA regularly but with substantial variation in cadence, social configuration, and interaction quality. Below, we summarize results from our data sources with respect to (1) engagement and use, (2) children’s experience and perceptions of content, (3) learning outcomes and interactional language patterns, and (4) breakdowns observed in real homes.

\subsubsection{Engagement and Use Over Time}
Children interacted with ELLA on average for $5.9$ out of $8$ days ($SD=2.02$, $min=2$, $max=8$, see Fig.~\ref{fig:ella-usage}). Parents reported time constraints as the primary reason for missed engagement. Across the deployment, each child engaged with $13.9$ stories on average ($SD=6.95$, $min=4$, $max=30$), corresponding to $2.56$ stories per active day on average. Individual story sessions lasted $5.80$ minutes on average ($SD=0.89$). Overall, the children engaged with ELLA for $14.01$ hours, averaging $1.41$ hours per child ($SD=1.11$, $min=0.50$, $max=2.72$). These patterns indicate consistent but flexible use, with meaningful household-to-household differences in frequency and intensity (Fig.~\ref{fig:ella-usage}). 

Based on parent diaries, most sessions occurred with parents and/or siblings present (Table~\ref{tab:initiation-presence-feeling}). Initiation also varied: some children primarily initiated sessions themselves (e.g., Isabella), others showed mixed child- and parent-initiated sessions (e.g., Sarah), and others were mostly parent-initiated (e.g., Natalie).

\subsubsection{Children’s Overall Experience with ELLA}
Children’s reported affect toward ELLA was broadly positive. Eight out of nine children indicated that they like or love ELLA ($M=4.1$ out of $5$; Fig.~\ref{fig:survey-results}) and similarly like or love talking to ELLA ($M=3.9$ out of $5$; Fig.~\ref{fig:survey-results}). One child (Andrew) reported that he did not like ELLA and did not like talking to ELLA because \pquotes{``she doesn’t want to talk to me.''} Andrew and his parent explained that this perception stemmed from ELLA failing to provide an additional story upon request one day during the deployment (due to a speech recognition error based on logs) and only having the goodbye story on the offboarding day. Parent diaries further characterized children’s in-the-moment feelings during sessions as most often neutral or excited, with some instances of frustration (mainly tied to interaction breakdowns and, in two cases, disinterest that day) and hesitation (Table~\ref{tab:initiation-presence-feeling}).

Children’s perceptions of being understood by ELLA were more mixed. Four children reported that ELLA understood them very well ($5$ out of $5$) and one child reported that ELLA understood them ($4$ out of $5$), while others reported feeling only partially understood ($n=2$, $3$ out of $5$) or not understood at all ($n=1$, Andrew, $1$ out of $5$). Children who reported partial or low understanding cited specific instances in which the robot misunderstood their intentions during story retelling or when requesting another story.

\subsubsection{Perceptions of Storytelling Content}
Across the deployment, ELLA generated $320$ stories spanning $71$ unique themes and targeting $32$ distinct vocabulary words (see Table \ref{tab:target-words-themes}). Each story included an average of $7.15$ instances of the target word ($SD=2.50$), providing repeated exposure within each session. Children generally evaluated ELLA’s stories positively ($M=4.2$ out of $5$; Fig.~\ref{fig:survey-results}), with two notable exceptions. One child rated stories as ``okay'' despite engaging with nearly all available stories, while another expressed dissatisfaction due to limited access to a small set of highly preferred themes (Tom and Jerry, Mickey Mouse, and Teenage Mutant Ninja Turtles).

\subsubsection{Vocabulary Learning Outcomes}
Results from the PPVT-style assessment indicate that interacting with ELLA supported vocabulary acquisition. On average, each child correctly recognized $2.8$ words on the post-deployment PPVT-style assessment ($SD=1.23$, $\mathit{Mdn}=3$). Children learned target words during the eight-day deployment. A one-way Wilcoxon signed-rank test showed the difference in the number of target words children knew from pre- to post-deployment ($M=2.6$, $SD=1.26$, $\mathit{Mdn}=3$) is significantly greater than zero ($z=27.5$, $p=.001$).

\subsubsection{Expressive Language and Interaction Strategies}
During interactions, ELLA employed multiple follow-up strategies to scaffold participation, including reducing choices ($364$ instances), extensions ($200$), co-participation ($86$), and eliciting responses ($32$). Moreover, children were observed using the target words when responding to ELLA (see Table \ref{tab:target-word-usage}). Children used more words when responding to ELLA each turn during the second half of deployment compared to the first half of the deployment. A one-way Wilcoxon signed-rank test showed that the difference in the average number of words the child said per turn is between days 1--4 and 5--8 ($M=0.92$, $SD=1.40$, $\mathit{Mdn}=0.57$) is significantly greater than zero (see Table \ref{tab:target-word-usage}), ($z=18.00$, $p=.037$).

Parents consistently reported that children adapted their speech in response to ELLA’s conversational constraints: children slowed down, spoke louder, and articulated more clearly to help the robot ``understand'' them. Several parents also noted that children made deliberate efforts to use longer, more complete sentences, particularly after being told during onboarding that full sentences would help ELLA respond appropriately. In some households, parents initially modeled or coached full-sentence responses and later observed children producing longer utterances independently.

\subsubsection{Interaction Breakdowns in the Home}
Speech recognition errors (\eg failure to hear, transcription mistakes) frequently manifested as interaction breakdowns. These included ELLA (a) proceeding to farewell and shut down even though the child requested another story, (b) providing feedback implying the child was wrong when the child provided the correct answer, and (c) missing responses. Based on parent accounts and interaction logs/videos, these errors were especially common in noisy environments (\eg background noise; other household members speaking), as ELLA could not distinguish speakers. In response, families adapted the environment and setup (\eg moving ELLA from a shared living room to a bedroom; hushing siblings; asking others to play elsewhere). Additional breakdowns observed in videos included (a) the robot interrupting the child, (b) difficulty handling child-initiated interruptions, and (c) responding too late to sustain conversational momentum.

Together, these results show that while ELLA supported vocabulary learning, elicited expressive language use and was generally well received, engagement and perceived interaction quality were shaped by real-world household conditions, multi-party participation, and the system’s ability to robustly support smooth conversational flow. We synthesize our broader findings into lessons learned and design opportunities in Section \ref{sec:lessons}.

\begin{table}[t]
    \centering
    \caption{Lessons Learned and Design Opportunities for ELLA}
    \label{tab:ella_lessons_dos}
    \scriptsize
    \setlength{\tabcolsep}{4pt}
    \renewcommand{\arraystretch}{1.05}
    \begin{tabular}{p{0.22\linewidth}p{0.74\linewidth}}
        
        \multicolumn{2}{l}{\textbf{Lesson:} Inviting Caregiver Partnership Could Enrich Language Development} \\
        \midrule
        \textbf{\colorbox{lightyellow}{Design Opportunity 1}} & Make learning gains visible through caregiver-facing reflective summaries. \\
        \textbf{\colorbox{lightyellow}{Design Opportunity 2}} & Extend language learning beyond session through robot-caregiver collaboration. \\
        
        \midrule
        \multicolumn{2}{l}{\textbf{Lesson:} Stories Need Freshness to Retain Playfulness Over Instructional Tone} \\
        \midrule
        \textbf{\colorbox{lightyellow}{Design Opportunity 3}} & Support low-effort, ongoing co-authoring to keep content aligned with shifting interests and goals. \\
        \textbf{\colorbox{lightyellow}{Design Opportunity 4}} & Balance meaningful repetition and novelty through narrative variation. \\
        
        \midrule
        \multicolumn{2}{l}{\textbf{Lesson:} Making Talk Feel Safe Needs Legible, Low-Pressure Interaction} \\
        \midrule
        \textbf{\colorbox{lightyellow}{Design Opportunity 5}} & Scaffold question complexity and turn-taking to children’s moment-to-moment readiness. \\
        \textbf{\colorbox{lightyellow}{Design Opportunity 6}} & Explicitly build and support children’s mental models of robot state. \\
        
        \midrule
        \multicolumn{2}{l}{\textbf{Lesson:} Robots Could Treat the Living Space as a Pedagogical Resource} \\
        \midrule
        \textbf{\colorbox{lightyellow}{Design Opportunity 7}} & Leverage multimodal intelligence to support embodied questioning. \\
        \textbf{\colorbox{lightyellow}{Design Opportunity 8}} & 
        Turn multi-party presence into a learning catalyst. \\
        
        \midrule
        \multicolumn{2}{l}{\textbf{Lesson:} Rigid Interaction Obligations are Incompatible with Family Routines} \\
        \midrule
        \textbf{\colorbox{lightyellow}{Design Opportunity 9}} & Sustain learning through flexible cadence and fitting into family routines. \\
        \textbf{\colorbox{lightyellow}{Design Opportunity 10}} & Enable spontaneous and proactive engagement through placement in shared-spaces. \\
        
    \end{tabular}
\end{table}

\section{Lessons Learned and Design Opportunities}
\label{sec:lessons}

Our home deployment of ELLA revealed how a generative AI-powered social robot can fit into children’s everyday lives supporting language development through playful, conversational storytelling. While the deployment demonstrated ELLA's potential, it also surfaced key insights into the complex realities of home-based engagement with pre-school aged children. Here, we distill these insights into lessons learned that reflect what worked, what proved challenging, and where new design opportunities emerge for child-robot interaction in this domain (See Table \ref{tab:ella_lessons_dos} for overview).




\subsection{Inviting Caregiver Partnership Could Enrich Language Development}

\subsubsection{Parents' Perception of ELLA} 
Parents widely perceived ELLA as valuable for language development, particularly vocabulary acquisition. Caregiver accounts suggest that observing children interact with ELLA highlighted the pedagogical value of the interaction by offering insights into aspects of children's development that are typically hidden from view at home, as much language instruction and assessment occur in school or therapeutic settings. Sarah's parents illustrated this by sharing: \pquotes{``We don't see her at school, so we have no idea about her comprehension or listening... You could hear ELLA say, 'Great job, Sarah!', you could hear Sarah answer [ELLA's] question correctly. That was really nice as a parent to be like, 'oh, my child can listen and pay attention and answer questions too.'''}. 

ELLA's perceived value also extended beyond typical pedagogical outcomes into complementing more targeted language development interventions. For example, Helen's parents shared \pquotes{``[Helen] mentioned [ELLA] to her speech therapist because we are practicing re-telling stories, and [the therapist] was very interested and happy to see [Helen] practicing with ELLA''}. Together, these observations suggest that caregivers' perceived value of robot interaction is shaped not only by what children learned, but by how clearly that learning could be observed, interpreted, and situated within broader developmental contexts. This indicates a critical need for visualizing this progress for caregivers to engage them as partners in children's language development without requiring continuous direct observation.

\textit{Design Opportunity: \colorbox{lightyellow}{Make learning gains visible through caregiver-facing reflective summaries.}} Future systems can surface children’s otherwise ``invisible'' learning through brief caregiver-facing reflections that capture both process and progress. These reflections can act as windows into emerging competencies by summarizing observable skills (e.g., listening comprehension, story recall, expressive use) and grounding them in lightweight evidence (e.g., a successful retell moment or a tagged interaction snippet). Such summaries can help at-home robot practice complement and connect to existing interventions (\eg speech therapy, school lessons).

\begin{figure}[!t]
  \includegraphics[width=\textwidth]{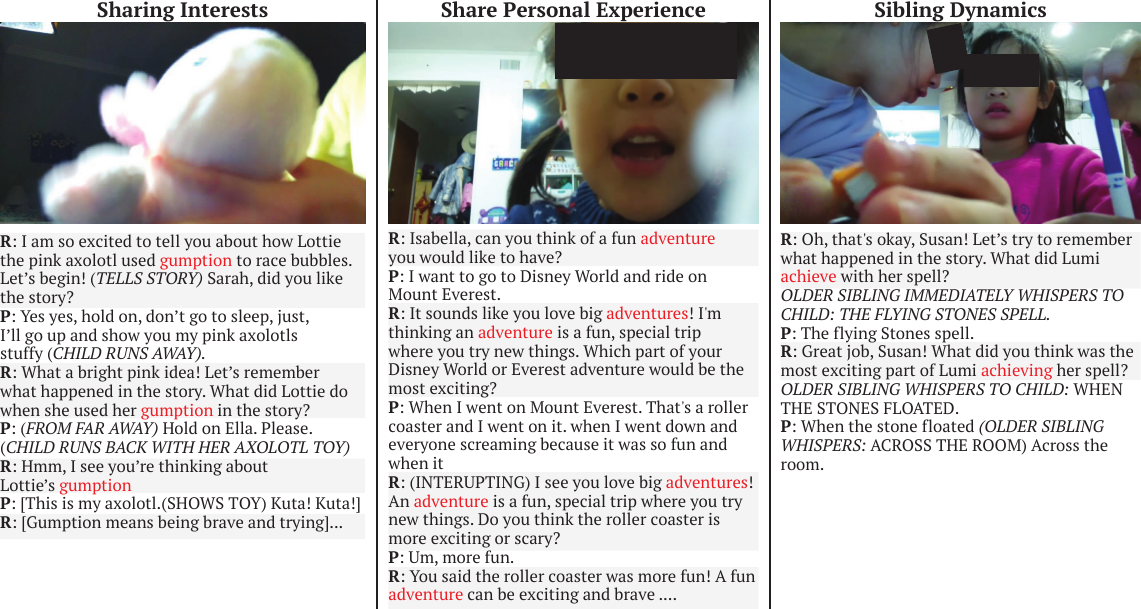}
  \caption{Our deployment data highlighted interesting interaction patterns that inspire new design opportunities. We highlight instances of object-based engagement, sibling dynamics, and sharing of personal experiences here. Pseudonyms are used, and the target vocabulary for the given interaction is highlighted in red.}
  \Description{This figure presents qualitative examples from the deployment that illustrate recurring interaction patterns with ELLA. The examples highlight three types of engagement: object-based engagement, sharing of personal experiences, and sibling dynamics.
  In the object-based engagement example, a child temporarily leaves the interaction to retrieve a personal toy related to the story theme and target vocabulary. In the personal experience example, the child connects the story to a real-life adventure and expands on their own experiences during the interaction. In the sibling dynamics example, an older sibling provides whispered assistance while the child responds to the robot’s questions.
  Pseudonyms are used for all children, and target vocabulary words referenced in the interactions are highlighted in red.}
  \label{fig:examples}
\end{figure}

\subsubsection{Language Use Beyond Story Session.} 
Children were observed using the target vocabulary both during and beyond story sessions, indicating that learning extended outside the immediate interaction. During interactions with ELLA, children applied target words not only within the narrative but also in novel ways, often relating them to personal experiences (see Fig. \ref{fig:examples}). Outside of story sessions, parents described children bringing target vocabulary into shared reading, play, and conversation, indicating transfer beyond the interaction itself.

For example, Sarah’s parent described a moment of parent-prompted vocabulary use during shared reading: \pquotes{``We were reading about a bike and a little girl who was trying again and again and again. I said, ‘Well, what is that?’ and [Sarah] goes, ‘Perseverance...and gumption!'''}. Parents also observed children using vocabulary spontaneously and testing its applicability in new situations. Jason’s parent described how their child applied target words during play while seeking confirmation of correct usage: \pquotes{``[when building a toy] he used ‘frisky’ to express his excitement and ‘clumsy’ when his toy broke into pieces since his hands weren’t controlling well.''}  

However, parents noted that transfer beyond the session depended on real-world applicability. Some target words (\eg Sheriff, Somersault) were described as infrequently used in the child's daily life, limiting opportunities for reinforcement outside of ELLA interactions. These observations suggest that children were not simply repeating words introduced by ELLA, but actively experimenting with their semantic fit across contexts—an indicator of deeper word learning. While some vocabulary naturally lent itself to everyday use, rarer or less familiar words required additional scaffolding. In these cases, parent prompting helped create opportunities for application, aligning with educator practices of sharing current learning topics with caregivers identified in our Phase 1 interviews.

Interestingly, this extension of language use was often intertwined with broader curiosity stemming from story themes. Parents described how stories sparked follow-up questions and exploration, such as Andrew asking his dad additional questions after a giraffe-themed story. Children’s curiosity also extended to the robot itself: Helen closely examined ELLA’s lights and asked how it worked; Susan wanted to make clothing for the robot; and Jason inspected the robot’s interior, noting that “it has a computer inside.” Together, these examples suggest that thematic curiosity motivated children to carry language, concepts, and inquiry beyond the story session and into everyday activity.

\textit{Design Opportunity:} \textit{\colorbox{lightyellow}{Extend language learning beyond session through robot-caregiver collaboration.}} Robots can treat caregivers as active partners in reinforcing and extending language practice beyond the session. Future systems could share short prompts or conversation hooks that help families weave target language concepts into everyday moments (e.g., bedtime reading, play, mealtimes). These prompts can also leverage children’s curiosity—for example, encouraging caregivers to explore related questions raised by stories (“What else do we know about giraffes?”) or by the robot itself (“How do you think the robot hears you?”). By embedding targets in familiar activities, robots can foster situated practice so language targets are reinforced in lived experiences beyond story sessions, thereby deepening engagement, enriching language development, and fostering habits of lifelong learning

\begin{figure}[t]
  \includegraphics[width=\textwidth]{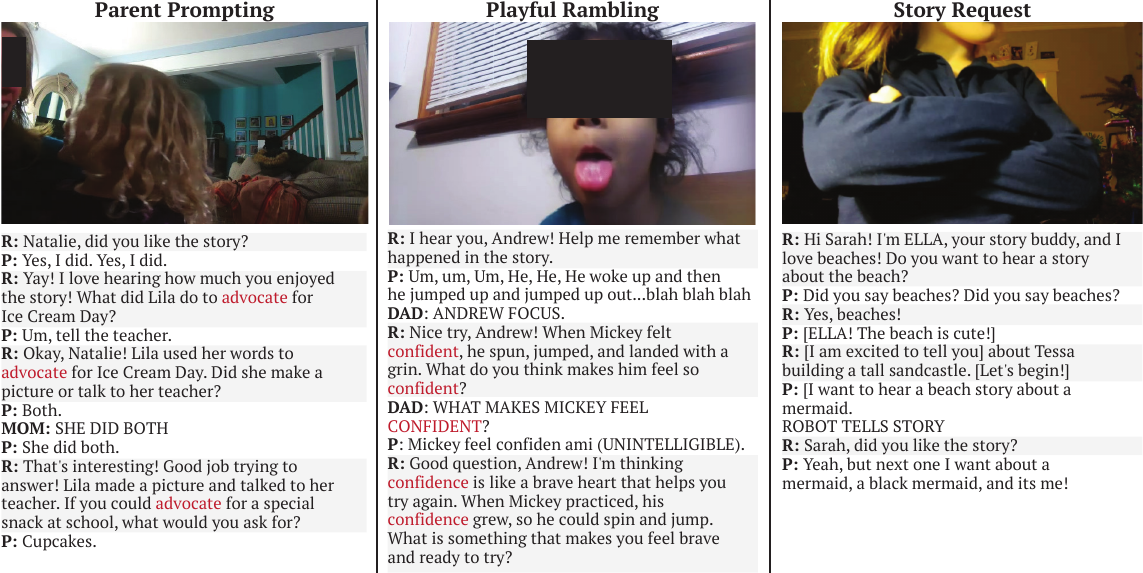}
  \caption{Examples demonstrating additional interaction patterns. We highlight instances of parent prompting, playful rambling, and story requests here. Pseudonyms are used, and the target word for the given interaction is highlighted in red. Overlapping speech is shown in brackets.}
  \Description{This figure presents examples from the deployment that illustrate recurring interaction patterns with ELLA. The examples highlight three types of engagement: parent prompting, playful rambling, and story request.
  In the parent prompting example, the parent helps model for the child how to say their one-word response in a full sentence. In the playful rambling example, the child responds with blah blah blah and unintelligible halfway through their sentence. In the story request example, a child interrupts Ella to ask for a specific story.}
  \label{fig:examples2}
\end{figure}

\subsection{Stories Need Freshness to Retain Playfulness Over Instructional Tone}

\subsubsection{Narrative Content.} 
Story engagement depended on how well stories tracked children’s current interests, which parents described as shifting day to day. When stories matched what children cared about, the experience felt playful and personally meaningful rather than instructional. Children actively expressed preferences and evaluated content. For instance, Sarah requested highly specific themes (see Fig. \ref{fig:examples2}) and voiced disappointment when stories missed the mark. Andrew similarly wished for more stories about the Teenage Mutant Ninja Turtles. These moments revealed that children were not passive consumers but intended to engage as curators of their narrative experiences. Parents’ reflections further suggested that story engagement was not only about themes, but also about the embedded learning goals. Several noted that some parent-selected target words, particularly abstract or decontextualized concepts (\eg Justice), were difficult for children to grasp. Together, these observations highlight a tension at the heart of generative storytelling for learning: maintaining pedagogical intent while preserving a sense of freshness, agency, and play.

\textit{Design Opportunity: \colorbox{lightyellow}{Support low-effort, ongoing co-authoring to keep content aligned with shifting interests and goals.}} Future systems can sustain playfulness by treating families as ongoing co-authors of the story experience. Generative AI can support caregivers in selecting and refining learning targets by previewing developmental fit and everyday applicability (\eg “high-frequency daily-use words” versus more abstract concepts), making tradeoffs visible, and suggesting alternatives when a target is unlikely to transfer. In parallel, robots can offer low-effort, recurring mechanisms for updating story ingredients\cite{cao2026reframing}such as: periodic caregiver check-ins (“What is Sarah into this week?”), quick pick-lists for story themes, and child-facing choices during sessions (“Pick one: mermaids, superheroes, animals”, "What should Milo do next?")) \cite{ligthart2020design}. During interaction, robots can further adapt story length, pacing, and linguistic complexity based on observed comprehension and uptake, introducing simpler retells or more grounded examples when needed. By enabling continuous, collaborative, low-effort alignment of story content and learning goals, robots can preserve the novelty and agency that keep storytelling both playful and pedagogically meaningful.

\subsubsection{Repition and Novelty} A fundamental tension exists between repetition as a learning mechanism and novelty as a motivator. Parents expressed a desire for greater story availability and flexibility as some children wanted to continue the session even after exhausting the day’s available stories; however, “more stories” does not necessarily mean new stories as Isabella's parent explained: \pquotes{“You can just repeat the stories. Don’t have to have new stories… for kids at this age, they like to hear stories over and over.”} At the same time, repetition interacted with vocabulary learning in ways that shaped engagement over time. Once children became familiar with the target words and story structure, some parents noticed a slow decline in excitement, describing sessions as feeling more like a instructional task than a playful experience. Sustained engagement, therefore, appears to depend on maintaining a sense of freshness,not necessarily through entirely new stories, but through variation that preserves surprise, agency, and narrative momentum.

\textit{Design Opportunity: \colorbox{lightyellow}{Balance meaningful repetition and novelty through narrative variation.}} To balance repetition as a learning mechanism with novelty as a motivator, future systems can decouple learning targets from the storytelling form. Instead of only swapping surface themes, systems can sustain engagement through structural variation: shifting plot arcs (e.g., mystery, quest, obstacle-and-retry, moral dilemma), character roles, settings, humor styles, and interaction patterns (e.g., question types, pacing, and placement). This approach can reduce topic fatigue, maintain excitement without sacrificing pedagogical goals, and better fit diverse developmental needs: younger children may benefit from short, predictable formats, while more advanced children may seek deeper plots, richer detail, and more open-ended expressive opportunities.

\subsection{Making Talk Feel Safe Needs Legible, Low-Pressure Interaction}

\subsubsection{Asking Questions.} Parents described how question complexity directly influenced children’s ability to produce spoken responses. Questions focused on immediate story events (\eg ``what just happened'') were generally accessible and supported responsive participation. In contrast, questions that required recalling earlier narrative details or making more abstract inferences disrupted some children’s ability to respond fluently, limiting expressive output. As George's parent noted, questions that referenced earlier points in the story imposed recall demands that constrained George's verbal responses: \pquotes{``It’ll get to a point of the story and then it’ll go to another part… and then it’ll ask him a question that calls back to the earlier point, which I feel it’s a little difficult for his age to kind of recall the details like that.''} Similarly, Andrew’s parent observed that while factual questions supported engagement, more interpretive prompts reduced successful responding: \pquotes{``He [would get] a little silly at times and, like, rambled [when] he didn't know what to say [in response to a question]''}. 


Although ELLA’s interaction design included a range of scaffolding techniques, some parents felt that children still needed more supportive scaffolding around their natural response patterns. Interaction logs revealed that frequent speech recognition errors---compounded by household noise, multiple speakers, rambling responses (see Fig. \ref{fig:examples2}), and occasional disengagement---made LLM-driven scaffolding inconsistent in practice, so strategies did not always manifest as designed. Together, these observations point to the need to more explicitly align question demands and scaffolding strategies with children’s expressive, mulit-modal language formats (e.g., single words, fragments, silences, or playful rambling) and their moment-to-moment readiness.


\textit{Design Opportunity: \colorbox{lightyellow}{Scaffold question complexity and turn-taking to children’s moment-to-moment readiness.}} Future systems should adapt questioning strategies to children’s real-time interactional readiness rather than treating all prompts as equivalent. Multimodal language models could enable sensing signals lost in the text modality (e.g., pauses, gaze shifts, hesitation, movement, or playful behavior) and can help robots infer states such as thinking, uncertainty, disengagement, or readiness to respond, and dynamically adjust how questions are asked. For example, when recall or inferential demands appear too high, the robot might simplify the question, re-anchor it in the immediate story moment, provide a partial scaffold, or allow extra time before responding. Rather than interpreting silence, rambling, or interruptions as failures, robots can treat them as signals to modulate pacing, prompt structure, or turn-taking—supporting more fluid, developmentally appropriate conversational rhythms that align with the fluid, nonlinear, and multimodal ways young children naturally communicate.

\subsubsection{Children's Mental Models} Over time, children formed mental models of how ELLA interacts with some moving from being ``reserved'' to being more ``conversational'' towards the end. Small behavioral cues played an outsized role in maintaining the mental model. For instance, ELLA's head tilt that indicates its thinking (\ie it has registered a response and is processing) helped support turn handoff and maintain the interaction flow. When this cue broke, children’s confidence in the interaction visibly shifted. In sessions where Helen’s robot motors were not powered due to a loose cable connection, the tilting behavior did not occur, and Helen began repeating her responses multiple times, unsure whether the robot had heard her. Similarly, George's parents described how ELLA's response latency did not match George's conversational rhythm, leading to confusion and prompting him to look to caregivers for help: \pquotes{``[George] would respond and the delay between his response and her next move would kind of confuse him. he'd be like, 'huh? what's going on?' and [then he'd] look for us.''}. Critically, we found that children treated being understood or not as relational with speech recognition errors being experienced not as technical ``errors,'' but as social refusal manifesting into negatively impacted perception (\eg \pquotes{“she doesn’t want to talk to me”}).

\textit{Design Opportunity: \colorbox{lightyellow}{Explicitly build and support children’s mental models of robot state.}} Future systems should explicitly support mental model formation by making internal states continuously legible through redundant, resilient cues (e.g., motion, lights, short verbal acknowledgements like “I heard you,” and backchannel sounds). Robots should leverage caregivers’ informal roles as ``interaction translators'' to ease children into participation by inviting co-participation early on and gradually shifting toward independent use as children gain confidence. Critically, when breakdowns occur (e.g., delayed responses, speech recognition failures, uncertainty in sensing), the system should explicity frame them as situational and technical, not relational, by offering (e.g., “I’m thinking…,” or “I’m having trouble hearing with the noise”) brief, child-legible explanations to prevent children interpreting errors as social refusal. Together, these strategies help stabilize interactional expectations, prevent social misreadings, and build a more resilient foundation for communication and learning.

\subsection{Robots Could Treat the Living Space as a Pedagogical Resource}

\subsubsection{Questions Can Invite Playful Action}. 
Beyond story content, parents’ accounts and interaction logs highlighted how the type of questions ELLA asked influenced children’s engagement and sense-making during storytelling. Two question patterns stood out: prompts that invited children to “show” understanding through action, and questions that appeared illogical or misaligned with children’s everyday reasoning.

“Show me”–style prompts often elicited lively, embodied responses. For example, when ELLA prompted around sympathy, Helen responded by hugging her parent (see Fig. \ref{fig:teaser}). These prompts appeared to align well with how children naturally express understanding through their bodies and environment, supporting playful engagement. At the same time, such prompts raised practical constraints. During deployment, we manually removed demonstration requests such as “[Jason], can you show me how you would do a somersault [the target word] with your body?” to avoid encouraging unsafe behavior. Parents also identified questions that “didn’t quite make sense,” particularly when answer options conflicted with children’s everyday logic (e.g., “Helen, can you think of something that is legal to do at home?”). While no parents reported feeling uneasy with ELLA’s behavior, and we did not observe unsafe or inappropriate content in interaction logs, these moments suggest that question design plays a critical role in shaping whether children can meaningfully engage with and respond to prompts.

\textit{Design Opportunity: \colorbox{lightyellow}{Leverage multimodal intelligence to support embodied questioning.}} Future robots should move beyond purely verbal Q\&A by intentionally supporting question types that invite children to show understanding through action, objects, and play. To make these prompts meaningful and practical, systems could (1) ground prompts in the child’s environment (e.g., “Can you show me something in this room that is soft?”) and (2) include safety and feasibility constraints that steer away from risky demonstrations while preserving playfulness. Multimodal language models can help the robot acknowledge what children do and treat gestures, shown objects, and physical engagement as valid responses and use them  as springboards to deepen narrative sense-making and language development.

\subsubsection{Robots turn into social artifacts in homes} The social configuration and the degree of caregiver involvement varied across families, from tightly shared ``family story time'' to largely independent child–robot sessions, with many families describing a transition from co-present interactions to more dyadic use over time. Parents often played active roles and shaped interaction by modelling how to answer, encouraging the use of more complete sentences, and managing environmental factors such as background chatter or siblings' presence, and occasionally helped the robot recover from errors. For example, due to a speech recognition failure, ELLA proceeded to say farewell and shut down even though Helen had requested another story. Helen immediately turned to her parent, and the parent helped repair and recover the interaction by saying, \pquotes{she must be really tired}. Co-presence also influenced how children engaged with ELLA. Some children were more confident when grandparents or friends were present (\eg Natalie was \pquotes{``less reserved with her grandma''}), while others focused more when parents and siblings were absent (\eg Andrew “\pquotes{did better when his brothers weren’t around}”).

Many children integrated ELLA within their broader social world in varied and meaningful ways. Some were eager to talk about her and introduce her into their social world, while others treated access to ELLA as something to protect. Helen, for instance, ``\pquotes{[wanted] to make sure her brother didn't join, becoming almost protective... which doesn't happen unless she's interested in something.}'' while Sarah proudly talked about ELLA at school:``\pquotes{I have a talking robot in my house...that was a big thing [for ELLA to be] in \textbf{her} house... she mentioned it once or twice in her classroom. She wasn't sure people believed her.}''. Children even invited visitors to watch or participate. During a playdate, Natalie and her friend Max (pseudonym) engaged in a joint storytelling session where ELLA told a story with the target word \textit{``advocate''}. Later, when Max’s mother arrived, Natalie explained that they had been talking to ELLA, and Max added, “When you \textit{advocate} for someone, you sort of stick up for them. And you do what’s right.” Such moments suggest that shared interactions with ELLA can support conceptual and vocabulary uptake in peer settings. At the same time, multi-party presence could sometimes dilute learning. When older siblings were present, they occasionally answered questions on the child’s behalf, reducing opportunities for independent thinking and expression (see Fig. \ref{fig:examples}). These patterns highlight that ELLA rarely sat in a pure dyad, it existed as a social artifact inside a dynamic social scene that can either amplify or dilute language practice. 

\textit{Design Opportunity: \colorbox{lightyellow}{Turn multi-party presence into a learning catalyst}} Rather than treating parents, siblings, and visitors as edge cases to manage, future systems should plan for multi-party engagement as a core interaction mode—amplifying learning while making the robot feel more approachable, especially for younger children who may be uncertain at first. Instead of enforcing single-speaker dialogue, robots could recognize overlapping speech and co-present behavior and scaffold it into the experience (\eg inviting siblings or peers to answer together, gently orchestrating turn-taking, or prompting brief moments of joint reflection). Audience-aware behaviors (such as lightweight self-introductions to newcomers, short recaps when others join, or invitations for children to explain) can transform social spill-ins into opportunities for expressive language use, narrative sense-making, and social retelling. By channeling the natural social “spill-in” of home interactions, these moments can reinforce expressive language, narrative comprehension, and conceptual understanding.

\subsection{Rigid Interaction Obligations are Incompatible with Family Routines.}

\subsubsection{ELLA in Daily Routines.} 
Parents consistently described ELLA as a new addition to family routines rather than a replacement for existing activities. As a result, fitting ELLA into daily life sometimes felt effortful—especially on busy school days when schedules were already crowded \cite{cagiltay2022understanding}. Natalie’s parent described evenings filled with \pquotes{``getting home, having a quick dinner, going to swimming… and then it’s time for bed,''} where adding a story session introduced stress, particularly after missing a day. Over time, attempting to maintain a strict daily cadence sometimes shifted the experience from novelty to obligation. As Sarah’s parent observed, \pquotes{``when she was doing it every day, she got bored and she got frustrated.''} These accounts suggest that mandating consistent daily engagement can undermine motivation and willingness to participate. At the same time, ELLA occasionally replaced traditional story time on busy days for some families, with children (\eg Andrew, Grace) interacting with ELLA as part of their bedtime routine showcasing how ELLA complemented existing rhythms rather than competing with them.

\textit{Design Opportunity: \colorbox{lightyellow}{Sustain learning through flexible cadence and fitting into family routines.}} Rather than enforcing consistent daily use, future robots should be designed to fit flexibly into the realities of family routines \cite{cagiltay2020investigating, cagiltay2023family} and be resilient to routine disruption. Because children’s schedules and energy levels fluctuate—and missed days are common—systems should aim to support pedagogically meaningful exposure without requiring full-length sessions every day. This might include “quick” sessions on busy nights, longer story time modes when families have more bandwidth, and gentle recovery after missed days (\eg a short recap or re-entry story that reinforces prior learning without restarting the routine). By reducing the pressure of daily completion and designing for disruption, robots can help sustain engagement over time and allow learning habits to form organically rather than turning interaction into an obligation.

\subsubsection{Robot placement.} Where ELLA lived in the home strongly influenced how and when it was used. Most families placed the robot in shared spaces such as the living room, envisioning it as a communal activity and a visual reminder to engage. This placement supported more child-initiated use and group interactions, but also introduced challenges such as increased ambient noise, reduced speech recognition reliability, and greater risk of and parental anxiety around damage during everyday activities. Although parents were generally not concerned about privacy, placement in shared spaces occasionally raised questions about sensing boundaries.

Other families chose to store ELLA in a safe space (\eg parent’s bedroom) to protect it from damage. While this reduced exposure to noise and accidents, it also reduced spontaneous use and increased parental effort. Parents had to remember to bring the robot out and prompt interactions, sometimes forgetting altogether. Grace’s parent shared \pquotes{It was two days where I completely forgot about it} and reflected that leaving ELLA out might have allowed Grace to use the robot independently during the day, and reduced parental burden.

\textit{Design Opportunity: \colorbox{lightyellow}{Enable spontaneous and proactive engagement through placement in shared-spaces}} Future robots should be designed to comfortably reside in shared household spaces where they can serve as a visible, low-friction invitation to interact rather than requiring storage, setup, or deliberate retrieval. This requires form factors and interaction designs that tolerate the realities of home environments and respectful co-presence through legible sensing states and boundaries, so families feel comfortable leaving the robot out. When situated in shared spaces, robots could support proactive engagement through gentle, context-sensitive invitations (\eg after extended inactivity, or when a child approaches), lowering the activation energy for use without adding pressure or interrupting family life.

\section{Limitations and Future Work}

While our findings highlight the promise of social robots in this domain, several limitations point to important directions for future work. Our study involved a small number of families and a short, eight-day home deployment, limiting insight into longer-term engagement, habit formation, and evolving caregiver roles. Future work should examine longer deployments to understand sustained use and routine integration. We focused on short-term receptive vocabulary gains and qualitative indicators of transfer. We did not assess longer-term retention, broader language outcomes (e.g., narrative production), or compare against alternative interventions. Future studies should include delayed post-tests, richer expressive measures, and comparative baselines. Our findings may not generalize across families with different cultural practices, languages, or material constraints. Differences in time, space, routines, and access to learning resources, particularly in lower-income households, can shape feasibility and use. Future work should study more socioeconomically and culturally diverse families to understand equity and necessary design adaptations.



\section{Selection and Participation of Children}
This study was approved by our Institutional Review Board (IRB). Participant recruitment was conducted through physical and electronic flyers, posts to newsletters, and community group chats. Participants were selected based on inclusion criteria that included children between the ages of four and six years old. Fourteen children from thirteen families were recruited. All parents consented and all children assented to participate in the study. Upon completion of the study, families were compensated \$10 gift card for the interview, \$20 gift card for the workshop, and \$40 gift card for the deployment study. Children were compensated additionally with stickers, patches, and 3D printed toys. 

\begin{acks}
\textbf{Funding.} This work is in part supported by the Johns Hopkins Malone Center for Engineering in Healthcare and National Science Foundation award \#2143704. 
\textbf{Author CRediT}: Conceptualization \& Methodology (VNA, SC, CH); Visualization (SC, VNA); Investigation; Review \& Editing (all); Original Draft (VNA, SC); Funding \& Supervision (CH).
\textbf{AI Statement}. This paper has been proofread by a language model. The authors verified that the resulting content accurately reflects the original intent. 
\textbf{Research Assistance.} We would like to acknowledge William Song for their contribution to recruitment and robot building, Victoria Popoola for their contribution to robot building and partly to data transcription; Qian Wang for initial ELLA prompt exploration; and Yoonjae Kim for robot building. 
\end{acks}

\newpage

\bibliographystyle{ACM-Reference-Format}
\bibliography{references}

\appendix

\end{document}